\documentclass[manuscript, screen=false, review=false, anonymous=false]{acmart}

\usepackage[yyyymmdd]{datetime}

\usepackage{float}
\usepackage{tabularx}
\usepackage{dcolumn} 
\newcolumntype{d}[1]{D{.}{.}{#1}}
\usepackage{subcaption}

\usepackage{todonotes}
\let\xtodo\todo
\renewcommand{\todo}[1]{\xtodo[inline,color=green!50]{#1}}

\AtBeginDocument{%
  }

\setcopyright{acmcopyright}
\copyrightyear{2018}
\acmYear{2018}
\acmDOI{XXXXXXX.XXXXXXX}

\acmConference[Conference acronym 'XX]{Make sure to enter the correct
  conference title from your rights confirmation email}{June 03--05,
  2018}{Woodstock, NY}
\acmISBN{978-1-4503-XXXX-X/18/06}

\begin{document}

\title{EEG and Eye-Tracking Evidence That AI Disclosure Shapes Face Evaluation}


\settopmatter{authorsperrow=3}

\author{Teodora Mitrevska}
 \orcid{0009-0005-4263-2510}
\affiliation{%
  \institution{LMU Munich}
  \city{Munich}
  \country{Germany}
}
\email{teodora.mitrevska@ifi.lmu.de}

\author{Luise Donat}
 \orcid{0009-0006-9593-3568}
\affiliation{%
  \institution{LMU Munich}
  \city{Munich}
  \country{Germany}
}
\email{luise.donat@campus.lmu.de}

\author{Andreas Butz}
\orcid{0000-0002-9007-9888}
\affiliation{
  \institution{LMU Munich}
  \country{Germany}
}
\email{butz@ifi.lmu.de}

\author{Thomas Kosch}
\orcid{0000-0001-6300-9035}
\affiliation{
  \institution{HU Berlin}
  \city{Berlin}
  \country{Germany}}
\email{thomas.kosch@hu-berlin.de}

\author{Abdallah El Ali}
\orcid{0000-0002-9954-4088}
\affiliation{%
  \institution{Centrum Wiskunde \& Informatica}
  \city{Amsterdam}
  \country{Netherlands}}
\affiliation{%
    \institution{Utrecht University}
    \city{Utrecht}
    \country{Netherlands}}
\email{aea@cwi.nl}
\email{abdallah.el.ali@cwi.nl}

\author{Francesco Chiossi}
\orcid{0000-0003-2987-7634}
\affiliation{%
  \institution{\href{https://mdxcro.com/}{MDx}}
  \city{Barcelona}
  \country{Spain}
}
\affiliation{%
  \institution{LMU Munich}
  \city{Munich}
  \country{Germany}
}
\email{francesco.chiossi@mdxcro.com}

\renewcommand{\shortauthors}{Mitrevska et al.}

\begin{abstract}
AI-generated faces can be difficult to distinguish from real ones, leaving viewers to rely on source labels when judging an image. Yet prior work has made it difficult to separate the effects of what an image actually is from what viewers are told it is.
We validated faces as AI-generated or human in an online study ($N=169$), then crossed actual source (AI, human) with label (none, Made with AI, Made by a human) in a lab study ($N=30$), recording event-related potentials (ERPs) and gaze.
ERP responses were equivalent for AI-generated and real faces, but varied with the label: labels drew early attention (N2), while labels that conflicted with the face's actual source prompted re-evaluation of the face (P3). Affective processing and initial gaze orienting were unchanged, but labels altered visual exploration.
We provide a validated stimulus set and evidence that attributed origin shapes face processing, with implications for disclosure design.
\end{abstract}





\begin{teaserfigure}
  \includegraphics[width=\textwidth]{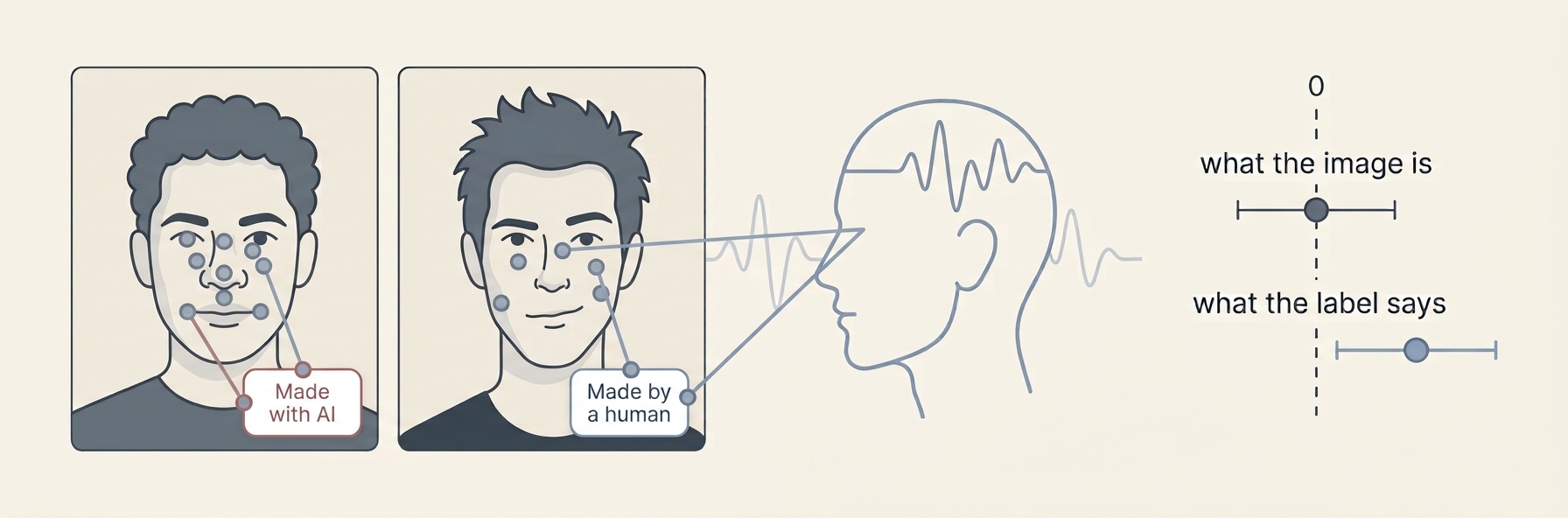}
  \caption{Observers cannot tell an AI-generated face from a real one, so the
  disclosure label attached to it becomes the only available cue to its origin. In the figure, two faces of identical apparent quality are depicted, one AI-generated and one real,
and are shown with the source labels used in our study. The scanpath illustrates how
participants moved between the face and the label while EEG and eye movements
were recorded. Faces carrying a label drew more and shorter fixations than
unlabelled ones. We found no clear difference between AI-generated and real
faces on any of our measures, while the origin attributed by the label was
reflected in how a face was processed. Image generated with Gemini Nanobanana 2.0 .}
  \Description{Two illustrated portraits, each carrying a small corner tag
  reading ``Made with AI'' or ``Made by a human''. Circles marking eye
  fixations are distributed across each face and on each tag, joined by lines
  into a scanpath that runs to the eye of a head shown in profile, with an EEG
  waveform drawn across the skull. At right, two rows of a forest plot share a
  vertical zero line: the row labelled ``what the image is'' has its point
  estimate on zero with the interval crossing it, and the row labelled ``what
  the label says'' has its point estimate to the right of zero with the interval
  clear of it. }
  \label{fig:teaser}
\end{teaserfigure}

\maketitle

\section{Introduction}

Generative models now produce synthetic faces that observers cannot reliably separate from
photographs of real people, and that are sometimes judged more trustworthy than genuine
ones~\cite{nightingale2022ai,lago2021more,tucciarelli2022realness,miller2023ai}. As
synthetic portraits spread across social platforms and online news
\cite{peng2025crafting}, the judgment of authenticity moves away from the pixels and
toward the cues that accompany an image. Source labels are one such cue, and disclosure
obligations for AI-generated content (cf.,~\cite{Elali2024aidisclosures}) are now written
into regulations across several jurisdictions~\cite{li2026governing}. Those rules require
only that a synthetic origin be disclosed. They take no position on how viewers process
that information.
This is relevant for how generated content should be disclosed. If a label only changes what
viewers report~\cite{feng2023examining, gangadharbatla2022role, drozd2026ai}, its effect
is available to them: they can weigh it, disagree with it, and set it aside if they
distrust the source. If a label changes how the face itself is seen, viewers cannot notice
that it did, so a wrong label produces a false impression they have no way to catch.
Whether labels reach face processing, or only the judgment that follows it, has not been
tested.

This distinction matters for how generated content should be disclosed. If a label only
changes what viewers report \cite{jakesch2019, drozd2026ai}, its effect is available to
them: they can inspect it and set it aside if they distrust the source.
If a label changes how the face itself is seen, viewers cannot notice that it did, so a
wrong label produces a false impression they have no way to catch. Whether labels reach
face processing, or only the judgment that follows it, has not been tested.


Two literatures approach the question from opposite sides. Work on synthetic faces changes
what an image is and leaves belief about its origin to chance. Work on labeling changes
belief, but does so over a single set of real photographs. In previous studies, actual
source and perceived source usually changed together, making it difficult to tell which
one caused the observed effect~\cite{miller2023ai, nightingale2022ai,
eiserbeck2023deepfake, conrad2025truthfulness}. Changing only the image cannot separate
the effect of the image from the effect of what viewers think it
is~\cite{miller2023ai, schindler2017}. Changing only the label keeps the actual
source of the face the same, so any difference between conditions could only show an
effect of the label, not whether responses also differ between AI-generated and human
faces. It also cannot test what happens when the label says that a face has a different
origin from its actual source~\cite{eiserbeck2023deepfake, conrad2025truthfulness,
jakesch2019}. Most of this evidence also rests on self-report, which records the outcome
of an evaluation and not the processing that produced it, and which is open to demand
characteristics when the manipulation is a visible label~\cite{gamage2025labeling}. The
few neurophysiological studies bring belief closer to perception, showing that believing a
face is fake dampens affective responses~\cite{eiserbeck2023deepfake} and narrows
gaze~\cite{cartella2024unveiling}.

We conducted two studies that separate a face's actual source from the source attributed to it. In Study~1, an online sample ($N = 169$) rated diffusion-generated faces for perceived AI-likeness, and we retained only images classified with high agreement. In Study~2 ($N = 30$), we crossed actual source (AI, Human) with label (No Label, Made with AI, Made by a human) within participants, recording EEG and gaze while each face was presented. Crossing the two factors produces congruent and incongruent pairings, so a response can be attributed to the label, the face, or the conflict between them. From the EEG, we analyzed three ERP components associated with different stages of face evaluation: the N2, a fronto-central marker of early mismatch detection~\cite{folstein2008, chiossi2024mind, kalamala2018flanker}; the P3, a centro-parietal index of stimulus re-evaluation~\cite{kao2025acute, yang2013does, chiossi2024understanding}; and the Late Positive Potential (LPP), a sustained signal of affective engagement~\cite{johnson2025association, ziereis2023motivated}.

AI-generated and human faces elicited the same N2, P3, and LPP amplitudes. Labeled faces
elicited a larger N2 than unlabeled faces, regardless of whether the label matched the
actual source. Faces whose label stated the wrong origin elicited a larger P3 than
unlabeled faces, consistent with that component's role in stimulus re-evaluation. The LPP
did not differ across any condition. Labels did not affect the latency of the first
fixation, but labeled faces received more fixations, and each fixation was shorter, than
unlabeled faces. The increase in fixation count was larger for AI-generated faces, and
fixations on AI-generated faces were shorter than on human faces regardless of label.
Actual source therefore affected how faces were visually explored even though it did not
affect any ERP component.

Taken together, these findings contribute in four ways. First, we show that realistic AI-generated faces did not produce clear differences from human faces in ERP responses or initial gaze orienting. Under these conditions, the face itself provided no reliable cue to its actual source, making the label an important source of information and leaving viewers with little basis for rejecting an inaccurate one. 
Second, we show that labels affect different stages of processing: their presence was reflected in the N2, while a mismatch between the label and actual source was reflected in the P3. Third, we also show that labels change visual exploration, producing more and shorter fixations without increasing total viewing time, which makes label placement a question of how attention is distributed. Finally, we open-source a validated stimulus set and design implications for disclosure labels, covering the accuracy required when viewers cannot check a label against the
content, the case for labeling human-made content alongside AI content, and the effect of
placement on how attention is divided between the disclosure and the image.

\section{Related Work}
Here, we review how observers perceive synthetic faces relative to real ones, covering realism and the uncanny valley, the finding that AI faces are frequently misidentified as human and can be judged more trustworthy than genuine ones, and the electrophysiological and evaluative signatures of believing a face is artificial.

\subsection{Perception of AI-Generated Faces}

Observers can no longer separate AI-generated portraits from photographs. In a large classification study, accuracy sat at chance, and a training session on rendering artifacts helped very little \cite{nightingale2022ai}. The same study found that generated faces were rated as more trustworthy than real ones~\cite{nightingale2022ai}, which is the opposite of what the uncanny valley predicts for faces this close to human~\cite{mori2012uncanny}, so a viewer's sense of unease is not a usable cue either. Faces of this kind are already in circulation. Yang and colleagues document a population of social media accounts that use them as profile pictures for scams and coordinated messaging~\cite{yang2024characteristics}.

Within a single generated set, though, some faces are much easier to place than others. \citet{miller2023ai} report that White AI faces were judged human more often than actual human faces, an effect they call AI hyperrealism, and that the participants who made the most errors were also the most confident. 
\citet{nightingale2022ai} see a similar racial difference in accuracy. The imbalance comes from the training data, so perceived race and perceived humanness are hard to disentangle, and Miller and colleagues advise checking that AI stimuli look equally natural across groups before running an experiment with them. The implication for stimulus design is that "AI-generated" is a fact about how an image was made, and it tells us very little about how any particular image will be seen.

A different line of work keeps the face fixed and changes what the viewer is told about it. \citet{eiserbeck2023deepfake} showed real photographs described either as authentic or as AI-generated. When a smile was believed to be fake, the usual emotional response to it largely disappeared from early visual and reflexive affective components, while later evaluative activity went up, which the authors read as more effortful appraisal. Angry faces were unaffected. A
belief about where an image came from can therefore reach perception, and it can do so early. \citet{mustafa2017human} find something comparable for virtual characters, where lower perceived humanness appears in components that index violated expectations about appearance and motion.
\citet{cartella2024unveiling} report that observers explore manipulated images
less widely than untouched ones, though they used natural scenes and not faces.

In both literatures the origin of a face and the origin attributed to it move together. Studies of the first kind change what the image is while leaving the viewer's belief to chance. Studies of the second change belief, but they do it over one set of real faces, so an effect could just as easily come from appearance. Telling the two apart needs images whose perceived origin is already known and agreed on. Study~1 produces such a set, and Study~2 crosses a label against it with appearance held constant.

\subsection{Effects of Content Labeling on User Judgments}
Source and provenance labels are the main platform-level response to synthetic
media. Credibility warnings on disputed content lower perceived accuracy and
reduce sharing intent, though the effects are modest and depend on how the
warning is worded and whether its source is
trusted~\cite{clayton2020, seo2019}. Labeling regimes also spill over: when only
some items carry a warning, the unlabeled ones gain credibility, an implied
truth effect that survives explaining the labeling policy to
participants~\cite{pennycook2020}. A label is therefore read together with the
content around it. For AI disclosure specifically, \citet{gamage2025labeling}
find that the wording, placement, and visual form of a label change what users
take it to mean.

A second line of work attaches a source to material of known origin.
\citet{jakesch2019} presented identical human-written Airbnb profiles as
AI-written or human-written, and found the AI attribution lowered trust only
when participants believed they were viewing a mixed set of both kinds, an
asymmetry they call the Replicant Effect. Disclosure of machine authorship also
depresses evaluations of visual art and poetry, even though participants
identify the true producer at near chance~\cite{ragot2020, kobis2021}. The MAIN
model explains this through cue-driven heuristics, where an interface cue
signaling machine origin activates a stored schema about machines that is
applied before any scrutiny of the artifact
itself~\cite{sundar2008, sundar2019}. A source label does not supply evidence
about the artifact; it supplies a category.

Two features of this work shape our design. First, the outcome is almost always a self-report collected after the stimulus has been processed, which cannot separate a label that changes what participants see from one that changes what they are willing to say. Second, actual source is held constant across conditions, usually because every item is human-made, so these designs estimate
a label effect but cannot estimate an effect of actual source, and cannot produce a trial in which the label contradicts the item. Crossing the two factors is what defines congruence, and measuring while the stimulus is on screen is what separates perception from report. Whether either point extends from text and artworks to faces is untested, and faces recruit dedicated perceptual machinery that the label must compete with.

\subsection{Electrophysiological and Eye Tracking Correlates of Face Processing}

The study of face perception has long relied on measures that do not depend on what a viewer chooses to report. Electroencephalography (EEG) records voltage changes at the scalp generated by populations of neurons. When that signal is
averaged across many trials time-locked to the same event, it yields event-related potentials (ERPs): a sequence of positive and negative deflections whose timing and scalp distribution have been mapped onto successive stages of processing.  Because a deflection at 170~ms and one at 600~ms correspond to different operations, the choice of which component to measure
determines what any effect can be said to demonstrate. Eye tracking provides a complementary record, capturing where the eyes move and how long they remain, and thus how attention is distributed across a display while it is being viewed.

Faces have a dedicated early signature. The N170 is a negative deflection around 170~ms that human faces evoke and other object categories do not, and it is taken to mark the point at which an image is encoded as a face~\cite{bentin1996}.
It arises in occipitotemporal cortex, so it appears at lateral posterior electrodes and cannot be recovered from central or parietal sites. The N170 also answers a question our design does not ask. Both AI and human portraits are
faces, and Study~1 has already fixed how they look, so what a source label plausibly affects is the evaluation of a face rather than its detection. We therefore measure three later components.

The first, the N2, is a negativity between 200 and 350~ms over frontal and central sites. \citet{folstein2008} separate a mismatch-related part of this component from a control-related one, both of which grow when a stimulus departs
from what the viewer expected. A label that contradicts the face it sits on
creates that kind of departure. 
The P3 comes next, a positive deflection over parietal sites that appears when a stimulus prompts the viewer to revise their current reading of the situation~\cite{polich2007}, which is what re-evaluating a face after registering a conflicting label would involve. The LPP continues that positivity for several hundred milliseconds and scales with how much the stimulus matters to the viewer~\cite{hajcakfoti2020}. Hajcak and Foti argue that the LPP and the P3 may reflect one underlying response to stimulus significance, so we read them as two windows on a single evaluative process. Earlier work on believed image origin reports effects in this late range, with enhanced late positivity for smiles that participants took to be synthetic~\cite{eiserbeck2023deepfake}.

Eye movements matter here for two reasons. The number and duration of fixations track how a viewer spreads and sustains processing across an image, and the latency of the first fixation captures initial orienting before any deliberate
strategy takes hold~\cite{rayner1998}. Beyond that, a label is itself a visible object occupying part of the display, so it can pull the eyes away from the face, and a difference in ERP amplitude between labeled and unlabeled trials could in principle follow from where participants looked rather than from what they understood the label to mean. Recording both lets us separate the two. The one comparable finding to date used manipulated scenes rather than faces, where viewers explored altered images less widely than untouched ones~\cite{cartella2024unveiling}, leaving open how gaze behaves when a disclosure label is attached to a face.

\section{Study 1: AI-Images Dataset Validation}
\label{study1}
The aim of Study~1 was to create a perceptually validated dataset of AI-generated facial images that human observers consistently recognize as synthetic. Although modern diffusion models such as Stable Diffusion can produce highly realistic portraits, the perceptual boundary between real and generated faces remains variable across outputs. For the purposes of Study~2, see \autoref{study2}, where we compare cognitive and multimodal responses to AI- and human-labeled images—it was essential to ensure that the stimuli labeled as ``AI-generated'' were not only technically artificial but also perceived as such by users.

To this end, we generated a diverse set of synthetic faces using Stable Diffusion and asked participants to classify whether each face appeared AI-generated or human. This rating task allowed us to identify a subset of faces that are reliably judged as artificial, ensuring perceptual consistency across experimental conditions in the main study.

\subsection{Procedure}
\label{procedure_study1}
The participants were first presented with an informed consent form where they were familiarized with purpose of this research, procedure, potential risks and rights associated with their data. After agreeing with the given terms, a demographics questionnaire was presented, collecting information about their age, gender and country of residence. As next, the participants were given a brief description of the task and a comprehension check. They proceeded with the task, where they were presented with an image of a human face. For each image, they had to express their judgment of the following statement "This image was created with the help of generative AI". Their judgment was collected via moving slider on a continuous scale, holding two end values : \textit{Strongly Disagree (left end value)} and \textit{Strongly agree (right end value)}. Additionally they expressed how confident they are in their judgment by using an exact same scale, following \citet{bray2023testing}'s approach. In the back-end the scores translated in numeric values where using the left end value (\textit{"Strongly disagree that the image is created with the help of generative AI"}) translates into 0 = definitely human, and using the right end value (\textit{"Strongly agree that the image is created with the help of generative AI"}) translates into 100 = definitely AI.

\subsection{Stimuli Generation}\label{stimuli_generation}

\begin{figure}[tbp]
  \centering
  \includegraphics[width=.85\linewidth]{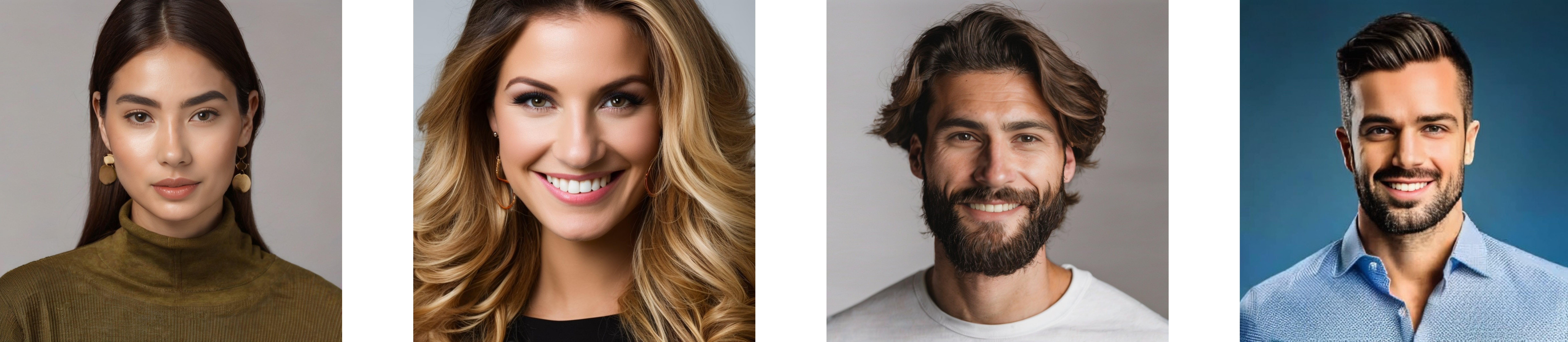}
  \caption{%
  Example images of AI-generated faces used in Study 1.}
  \label{fig:example_images}
\end{figure}

We chose the  initial pool of 320 facial images from the standardized \textit{Stable Diffusion Face Dataset} \footnote{https://github.com/tobecwb/stable-diffusion-face-dataset?tab=readme-ov-file}, generated via Stable Diffusion 1.5, 2.1, and SDXL 1.0. 
The images were generated using simple prompts such as \textit{"photo of a woman"} and underwent a quality control check as well as additional editing to fix any skewed facial features by the author of the database. 
All images have a resolution of 512 × 512 pixels and are subsequently inspected for artifacts (e.g., facial asymmetries, anatomical distortions). This curated image set served as the input for our validation task in  Study 1. An example subset is depicted in \autoref{fig:example_images}.

\subsection{Sample Size Determination}

Sample size planning was guided by the goal of achieving adequate precision for continuous 0--100 AI-likeness ratings \cite{imbault2018reliability,ebner2010faces}. Each image was to be classified as ``reliably AI-looking'' when the lower bound of the 95\% confidence interval around its mean rating exceeded 60 points \cite{winkler2009properties}. To ensure that confidence intervals were sufficiently narrow not to cross this threshold, we targeted an average margin of error of approximately $E=6$ points on the rating scale, consistent with practical constraints on participant load.

Assuming a standard deviation of $\sigma=20$ points, consistent with prior perceptual rating studies \cite{ebner2010faces}, the required number of ratings per image was calculated as:
\begin{equation}
n = \left(\frac{t_{\alpha/2} \times \sigma}{E}\right)^2 = \left(\frac{1.96 \times 20}{6}\right)^2 \approx 40.
\end{equation}

Thus, approximately 40 ratings per image were required to achieve the desired level of precision. Given 180 total images and 40 ratings per participant, we estimated that recruiting around 175 participants would provide complete stimulus coverage while maintaining independent observations across raters. The final dataset included 169 participants after quality control, yielding 6{,}760 valid trials and meeting the planned level of precision for reliable image classification. Empirically, the average half-width of the 95\% confidence interval across images was 4.8 rating points, confirming that the achieved precision was within the targeted margin of error.

\subsection{Participants}

We recruited 178 participants (\(N = 178\)) through Prolific Academic.  
The sample included 97 men (54.5\%), 77 women (43.3\%), 3 non-binary participants (1.7\%), and 1 participant who self-described their gender.\footnote{Percentages may exceed 100\% due to rounding.}  
Participants’ ages ranged from 20 to 78 years (\(M = 37.49\), \(SD = 12.45\)).  
The age distribution was broadly representative of an adult population, with 18–24 years (14.7\%), 25–34 years (40.1\%), 35–44 years (18.6\%), 45–54 years (16.5\%), 55–64 years (7.1\%), and 65+ years (2.9\%).  
Most participants reported English (\textit{EN}) as their survey language (99.7\%), and the most represented countries were Australia (10.9\%), Canada (8.2\%), the United States (7.6\%), South Africa (5.3\%), and the United Kingdom (5.0\%).  

After applying quality-control criteria (see \autoref{study1_dataProcessing}), where we we excluded nine speeders (median per-trial time \(< 4.45~\text{s}\))—169 participants (\(N = 169\)) and 6{,}760 valid trials were retained for analysis.

\subsection{Data Processing}
\label{study1_dataProcessing}

All preprocessing and analyses were conducted in Python. Trials with missing or non-numeric values in \texttt{Rating} or \texttt{Time} were discarded. To ensure data quality, we implemented a multi-step filtering procedure combining trial-level and participant-level checks. At the trial level, univariate outliers were identified and removed for \texttt{Time}, \texttt{Confidence}, and \texttt{Rating} using an interquartile-range criterion (\(1.5\times IQR\) beyond the 25th or 75th percentile). At the participant level, we excluded individuals who showed atypically short completion times (``speeders,'' defined as a median per-trial time below the 5th percentile of the group distribution, corresponding to \(< 4.45~\text{s}\)), participants with very low variability in responses (``flatliners,'' defined as having both rating SD and IQR below 5), and participants contributing fewer than 20 valid trials. This filtering procedure excluded nine participants (all classified as speeders), resulting in 169 retained participants out of 178 total (\(95\%\)) and 6{,}760 valid trials out of 7{,}120 total (\(95\%\)). 


\subsection{Analysis}
After collecting the ratings described in \autoref{procedure_study1}, we computed the mean rating for each image, standard deviation, and 95\% confidence interval (CI) across all valid responses. 
Images were classified as \emph{AI-looking} when the lower bound of the 95\% CI exceeded 60 points, ensuring that even the most conservative estimate of perceived ``AI-likeness'' remained above the neutral midpoint. 

To further characterize perceptual agreement, we calculated (a) the proportion of raters per image whose individual rating exceeded 70 on the 0--100 scale, (b) standard deviations across images as a measure of within-image rating variability, and (c) internal consistency and split-half reliability across participants. 
Internal consistency was estimated using Cronbach’s $\alpha$, and reliability between independent rater halves was assessed with a split-half correlation ($r_{\mathrm{split}}$) and its Spearman–Brown–corrected value ($r_{\mathrm{SB}}$).

\subsection{Results}

\begin{figure}[tbp]
  \centering
  \includegraphics[width=.7\textwidth]{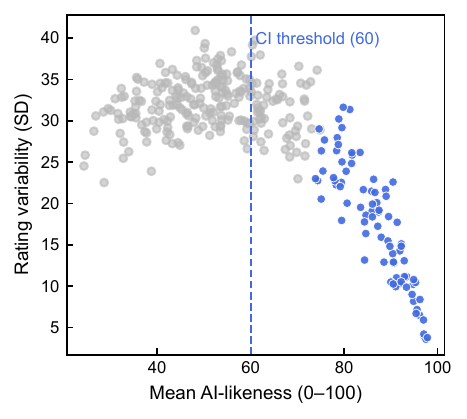}
  \caption{\textbf{Perceived AI-likeness vs.\ rating variability.} Relationship between mean AI-likeness rating and rating variability (SD) across the 320 AI-generated faces. Blue points mark the 82 images whose lower 95\% CI $\geq 60$, satisfying the inclusion criterion for the validated set. The dashed line indicates the CI-based classification boundary.}
  \label{fig:study1_scatter}
\end{figure}

\begin{figure}[tbp]
  \centering
  \begin{subfigure}[t]{0.47\linewidth}
    \centering
    \includegraphics[width=\linewidth]{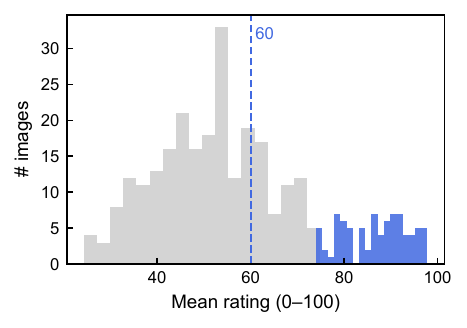}
    \caption{Distribution of mean ratings across all 320 AI-generated faces. Blue bars correspond to the 82 images retained as reliably AI-looking (lower 95\% CI $\geq 60$); gray bars represent discarded or ambiguous items. The dashed line marks the perceptual threshold at 60 points.}
    \label{fig:study1_hist}
  \end{subfigure}\hfill
  \begin{subfigure}[t]{0.47\linewidth}
    \centering
    \includegraphics[width=\linewidth]{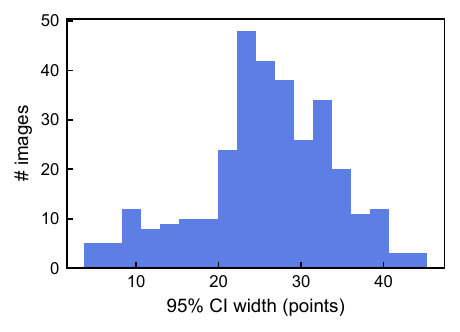}
    \caption{Distribution of 95\% confidence-interval widths across the 320 AI-generated faces. The narrow spread (median $\approx 9$ points) indicates consistent and precise judgments of AI-likeness.}
    \label{fig:study1_ciwidth}
  \end{subfigure}

  \caption{\textbf{Summary of perceptual consistency metrics.} (a) shows how mean ratings distribute around the inclusion threshold; (b) illustrates the achieved precision of participants' judgments via CI widths.}
  \label{fig:study1_hist_ci}
\end{figure}

Applying the CI-based classification criterion (Figure~\ref{fig:study1_scatter}), 82 of the
320 synthetic faces (25.6\%) were identified as reliably AI-looking, whereas the remaining
238 were perceived as human-like or ambiguous and excluded from the validated stimulus pool
used in Study~2. 
As shown in Figure~\ref{fig:study1_hist}, the distribution of mean ratings exhibits a clear separation between reliably AI-looking and discarded images. 
Across AI-looking faces, the average rating was $M = 78.4$, $SD = 6.5$, and the mean width of the 95\% CI was 9.2 points (Figure~\ref{fig:study1_ciwidth}), 
indicating strong perceptual convergence. 
Images below the 60-point threshold had lower mean ratings ($M = 48.1$, $SD = 10.3$) and wider confidence intervals (mean width = 14.5 points).

On average, 84\% ($SD = 7.2\%$) of participants rated the AI-looking images above 70, 
demonstrating strong consensus that these faces appeared artificial. 
For non–AI-looking images, the corresponding proportion was 41\% ($SD = 12.5\%$). 
Internal consistency across raters was high (Cronbach’s $\alpha = .90$), 
and a split-half reliability analysis (randomly dividing raters into two groups; $r_{\mathrm{split}} = .70$, Spearman–Brown corrected $r_{\mathrm{SB}} = .83$) 
confirmed stable perceptual judgments across independent subsets of participants.

\begin{figure}[tbp]
  \centering
  \includegraphics[width=.85\linewidth]{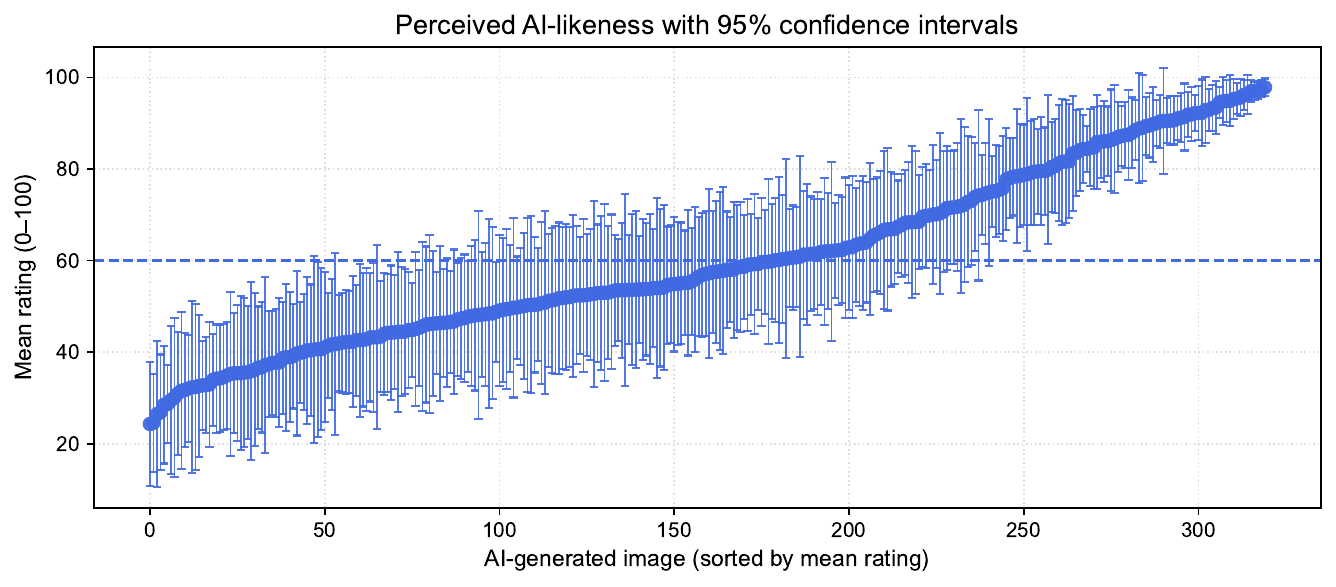}
  \caption{\textbf{Sorted mean ratings with 95\% confidence intervals.} Mean AI-likeness ratings for all 320 AI-generated faces, ordered by mean value. Error bars represent the 95\% CI per image, and the dashed line at 60 marks the classification boundary used to identify the 82 reliably AI-looking stimuli.}
  \label{fig:study1_errorbars}
\end{figure}

The overall distribution of image means and confidence intervals is illustrated in Figure~\ref{fig:study1_errorbars}, 
showing a distinct cluster of faces with consistently high AI-likeness ratings above the 60-point boundary. 
Together, these results indicate that the final set of 82 AI-generated faces were consistently perceived as artificial with high confidence and cross-rater agreement, 
providing a validated and reliable stimulus base for the subsequent experimental study.

\subsection{Discussion and Conclusion}

Study~1 validated a perceptually consistent set of AI-generated facial images that participants reliably identified as synthetic. 
The results demonstrate that perceptual judgments of “AI-likeness’’ are stable across raters and largely independent of individual noise, as indicated by high internal consistency and narrow confidence intervals. 
This validation provides an empirically grounded method for separating convincingly artificial from ambiguous faces—an important prerequisite for investigating cognitive and affective reactions to AI-generated imagery.

By isolating 82 faces whose lower 95\% confidence bound exceeded 60 on the AI-likeness
scale, we ensured that the subsequent experiment would manipulate perceived image source
without confounding effects of perceptual uncertainty. The criterion was deliberately
conservative, retaining roughly a quarter of the initial pool, which reflects how variable
the perceptual status of diffusion-generated portraits remains across outputs.
These validated stimuli form the basis for Study~2, where we examine how explicit source labeling (AI- vs.\ human-generated) influences participants’ cognitive processing and physiological responses.

\section{Study 2: Electrophysiological Responses to Labeled AI and Human Faces}
\label{study2}

Our study investigates how users perceive and evaluate faces that are either AI-generated or real, and how this perception is influenced by the presence and type of label. We presented participants with carefully controlled facial images and captured their behavioral responses and ERP signals to assess attentional and affective processing. 

Based on prior literature on visual cognition, trust, and electrophysiological responses to facial stimuli, we investigate the following research questions:

\begin{enumerate}

    \item[\textbf{RQ1}:] Does the actual source of a face (\textit{AI vs.\ Human})
  modulate its electrophysiological processing across early attentional (N2), evaluative
  (P3), and sustained affective (LPP) stages?

  \item[\textbf{RQ2}:] Does the \emph{congruence} between a face's source and
  its label---rather than the source itself---drive ERP responses, such that
  label--face conflict is reflected in components indexing conflict detection
  (N2) and stimulus re-evaluation (P3)?

  \item[\textbf{RQ3}:] Do disclosure labels alter how a face is visually
explored (the number and duration of fixations) and does this depend on the face's actual source?

\item[\textbf{RQ4}:] Does source or labeling affect the earliest allocation of attention, as indexed by the latency of the first fixation?

\end{enumerate}

\subsection{Study Design}

We employed a \(2 \times 3\) within-participant experimental design, with the independent variables \textsc{Source} (two levels: \textit{AI-generated}, \textit{Human}) and \textsc{Label} (three levels: \textit{No Label}, \textit{Made with AI}, \textit{Made by a human}). \textsc{Source} referred to the actual origin of the facial image, whereas \textsc{Label} indicated the presented information label (or absence thereof), delivered using minimalistic white rectangle holding the respective label.

To control for potential order and carryover effects, we implemented a balanced Latin square design across participants \cite{wang2009construction}. This procedure ensured that each of the six experimental conditions appeared equally often in each serial position and preceded and followed every other condition an equal number of times across the sample.



\subsection{Procedure}

Upon arrival, participants were welcomed into the lab, informed about the study protocol, and provided written informed consent. They then completed a short questionnaire including demographic information (age, gender, education level, and employment status) and an AI literacy assessment~\cite{carolus2023mails}. To ensure participants understood the task, we presented a brief explanation along with a few example trials on paper before beginning the setup.

Participants were fitted with a 64-channel EEG cap and seated in a comfortable, ergonomically adjustable chair at a fixed distance of 60~cm from the screen. They placed their chin and forehead on a headrest to minimize movement and ensure consistent viewing distance. An EyeLink 1000 Plus eye tracker was calibrated using a standard 9-point calibration procedure. 

The main task consisted of 240 trials, organized in a $2 \times 3$ within-subjects design manipulating the actual source of each face (AI-generated vs. Human) and the associated label (No Label, Made with AI, Made by a human).

\begin{figure}[tbp]
  \centering
  \includegraphics[width=.85\linewidth]{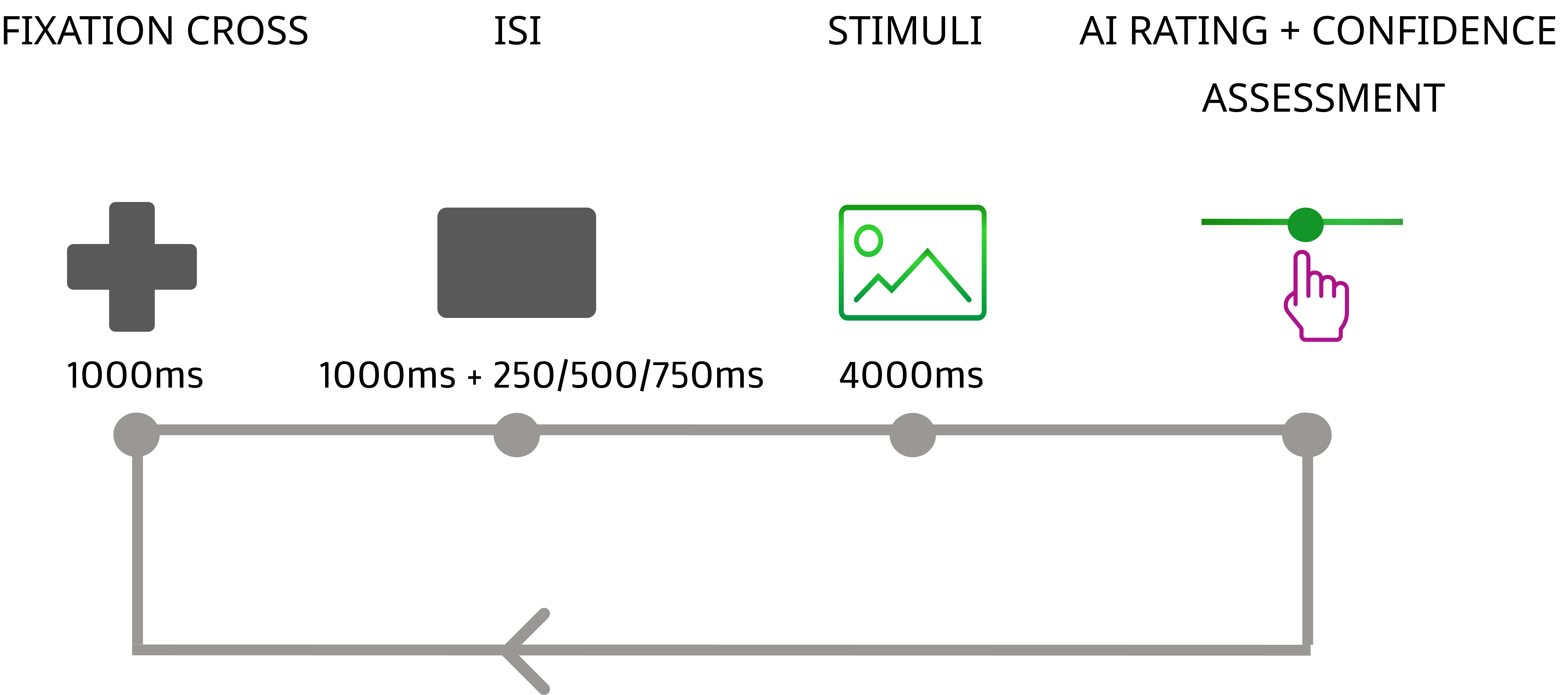}
  \caption{%
  Trial structure. Each trial began with a fixation cross (1000\,ms),
followed by a blank inter-stimulus interval (1000\,ms plus a random jitter of
250, 500, or 750\,ms) to reduce temporal expectancy and ERP overlap. A face was
then shown for 4000\,ms, either AI-generated or human and carrying a
\emph{Made with AI}, \emph{Made by a human}, or no label, while EEG and eye
movements were recorded. After the face offset, participants rated on a
continuous scale whether the image was created with the help of generative AI
and reported their confidence. Trials followed one another in fully randomized
order.}
  \label{fig:trial_structure}
\end{figure}

We structured each trial in line with established ERP protocols for face evaluation studies, particularly following the approach used by ~\citet{conrad2025truthfulness}. Each trial began with a black fixation cross presented for 1000 ms to stabilize gaze and establish a baseline. This was followed by a blank screen, i.e., Inter-Stimulus Interval (ISI), lasting 1000 ms plus a random jitter of 250, 500, or 750 ms to reduce temporal expectancy and overlap of ERP components. The stimulus image—either an AI-generated or real human face, with or without an accompanying label—was then shown at the center of the screen for 4000 ms. Condition order followed the balanced Latin square described above, with the
particular face assigned to each condition slot randomized within participant. Identically as in \autoref{procedure_study1}, after the image offset the participants used a continuous scale to  had to express their judgment on whether "This image was created with the help of generative AI", ranging values from : \textit{Strongly Disagree (left end value)} and \textit{Strongly agree (right end value)}. Additionally, they expressed how confident they are in their judgment, using an exact same scale. Trial timing and structure followed the sequence outlined in \autoref{fig:trial_structure}.
Participants could pause and take a break between each block to minimize fatigue and physical strain. The entire session, including setup, calibration, and task completion, lasted approximately 45 minutes.

\subsection{Stimuli}

\begin{figure}[tbp]
  \centering
  \includegraphics[width=.45\linewidth]{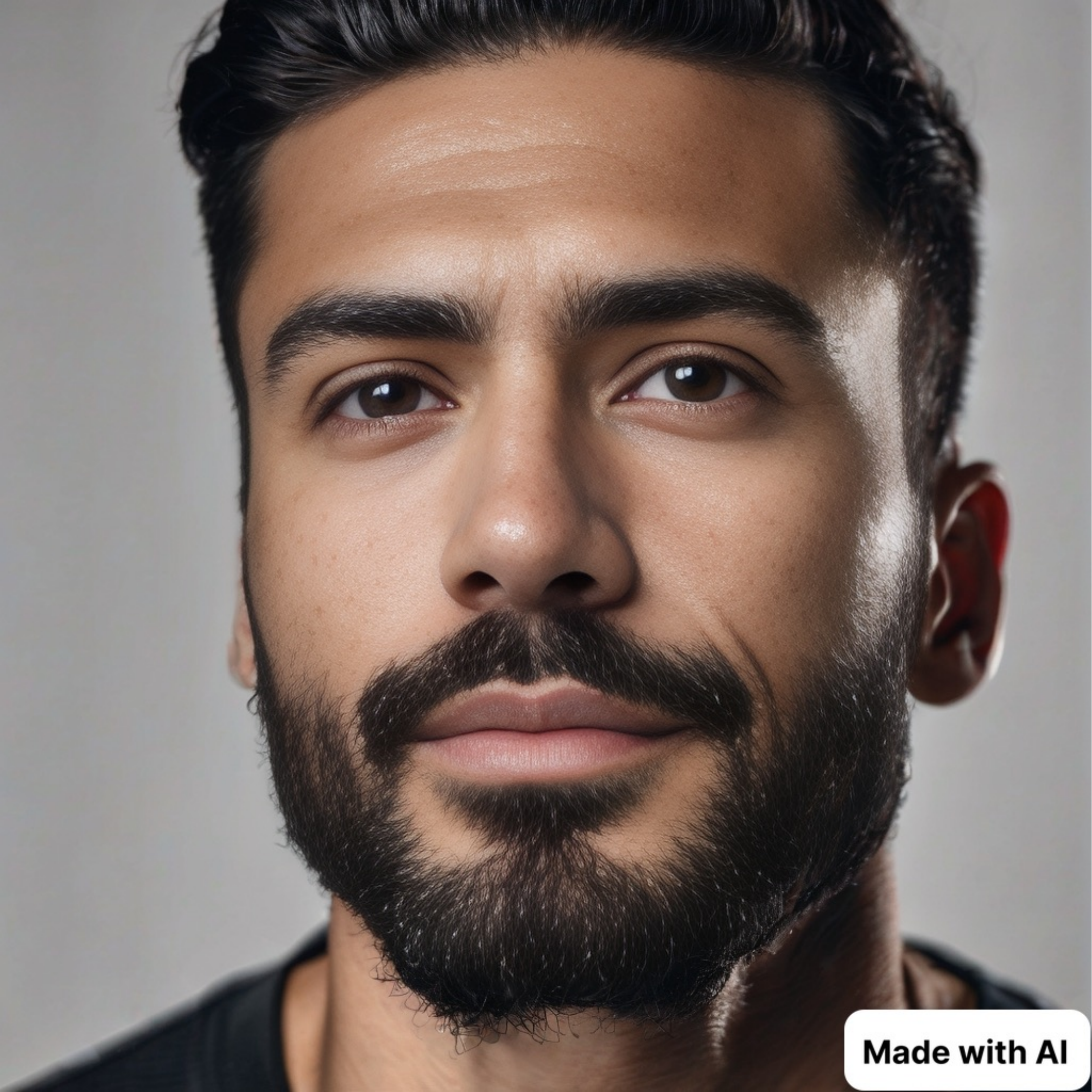}
  \caption{%
  An example of how the label was positioned in the bottom right corner, following \citet{gamage2025labeling}'s work.}


  \label{fig:label_example}
\end{figure}

The experimental stimuli consisted of 240 images, with 120 images in each of the two evenly distributed categories: AI-generated faces and real human faces. Each image was paired with one of three label conditions, i.e.,  "No label", “Made with AI” or “Made by a human”, yielding a fully crossed 2×3 design to systematically investigate how both the actual source and the assigned label influence perception.
We designed the image stimuli to control for extraneous factors known to influence ERP responses, such as brightness, color, and the presence of text, as established in prior EEG and ERP research \cite{eroglu2020brightness, anders2006erp, nissen2021gender}. The images of human faces were sourced  from the \citet{Ma2021Chicago}'s Chicago Face Database, which offers standardized stimuli with high image quality and ensures demographic diversity, encompassing representatives of multiple ethnic and racial groups. 
To convey the image source label (e.g., Made with AI or Made by a human), we used minimalist white rectangular box with the respective label and positioned it in the bottom right corner. This design decision follows \citet{gamage2025labeling}'s design guidelines (depicted in \autoref{fig:label_example}), suggesting that this label location provides a balance between visibility and minimal interference with image interpretation, while effectively communicating AI involvement in the created content.

\begin{figure}[tbp]
  \centering
  \includegraphics[width=.85\linewidth]{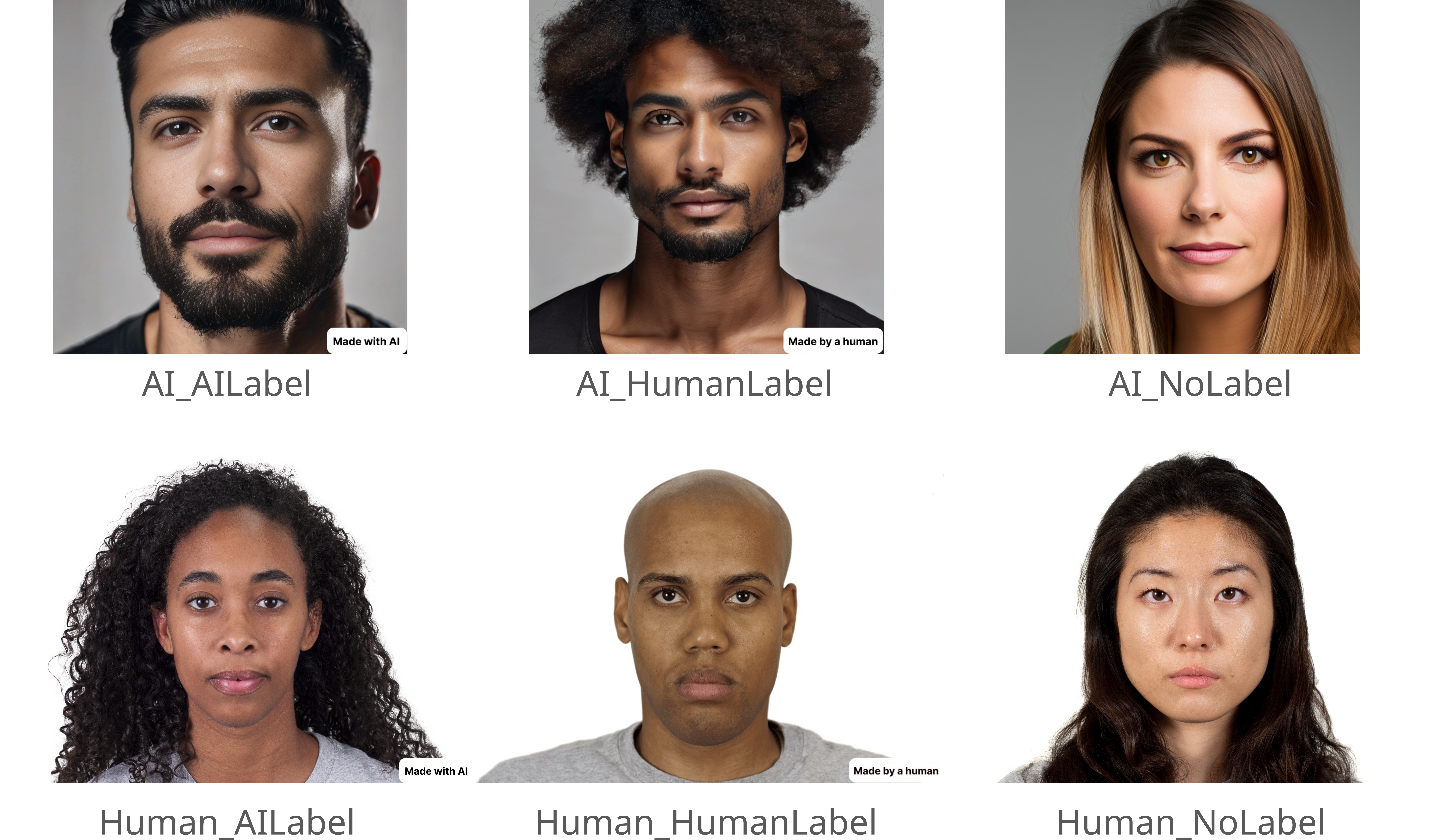}
  \caption{%
  Example images used in Study 2. In the first row, all images are AI-generated faces, representing all three label possibilities from left to right: Made with AI, Made by a human, No label. In the second row, all images are real humans, also representing the above mentioned labeling possibilities.}


  \label{fig:condition_images}
\end{figure}

\subsection{Apparatus}
The study was implemented in PsychoPy (v2025.2.0). Stimuli were presented on a VIEWPixx LCD monitor (23.6\,in, $1920\times1080$, 120\,Hz), with an active display area of $52.2\times29.4$\,cm and a pixel pitch of 0.272\,mm. Participants viewed the screen from a fixed distance of 60\,cm on a chin-and-forehead rest. Face images were
presented centrally at their native resolution of $512\times512$\,pixels, corresponding to $13.9\times13.9$\,cm and subtending $13.2^{\circ}$ of visual angle in both dimensions. The source label occupied the bottom-right corner of the image, placing its centre at approximately $7^{\circ}$ of radial eccentricity from fixation, outside the parafoveal region, so that resolving it required a saccade away from the centre of the face. Synchronisation of EEG, eye tracking, and experimental events was achieved with the Lab Streaming Layer (LSL) framework~\cite{kothe2025lab}.

\subsection{Eye Tracking Recording \& Preprocessing}

Binocular eye movements were recorded with a video-based EyeLink 1000 Plus (SR Research, Ottawa, Canada) at 1000~Hz, with the built-in heuristic filter. The head was stabilized on a chin-and-forehead rest at 60~cm. We calibrated with a 9-point procedure and validated immediately after, repeating until mean error was below $0.5^{\circ}$ and maximum error below $1.0^{\circ}$. Data were recorded and presented with EyeLink software (2.6.11 version), PsychoPy (v2025.2.0 version), and SR Research firmware (5.50 version). Fixations were parsed online by the EyeLink saccade detector (velocity $\geq 30^{\circ}/\mathrm{s}$ or acceleration $\geq 8000^{\circ}/\mathrm{s}^{2}$).\footnote{\url{https://natmeg.se/onewebmedia/EL1000_UserManual_1.52.pdf}}

Because recording was binocular, a single dominant eye was selected per participant for further fixation and saccade analysis, defined as whichever eye yielded the greater number of valid fixations across the recording session per participant.  For each trial, fixation and saccade events were retained if their onset fell within the stimulus presentation window (4000ms). First-fixation latency was defined as the onset of the first retained fixation relative to image onset. 
The region of interest (ROI) was defined as the full image region based on image origin : the images from the Chicago Face Database spanned 1067 x 750 px, and the Stable Diffusion Face Dataset spanned 750 x 750 px, centered within a 1920 x 1080 display. For each fixation, we computed the distance to the AOI and converted it to degrees of visual angle. We introduced a tolerance of $0.5^{\circ}$ visual angle, to include fixations that sat on the frame edge on the presented image.

\subsection{EEG Recording}
EEG data were collected using a 64-channel water-based R-Net cap (BrainProducts GmbH, Germany) with Ag-AgCl pin-type passive electrodes arranged according to the international 10--20 system. Electrodes were placed at the following locations: Fp1, Fz, F3, F7, F9, FC5, FC1, C3, T7, CP5, CP1, Pz, P3, P7, P9, O1, Oz, O2, P10, P8, P4, CP2, CP6, T8, C4, Cz, FC2, FC6, F10, F8, F4, Fp2, AF7, AF3, AFz, F1, F5, FT7, FC3, C1, C5, TP7, CP3, P1, P5, PO7, PO3, Iz, POz, PO4, PO8, P6, P2, CPz, CP4, TP8, C6, C2, FC4, FT8, F6, F2, AF4, and AF8. Signals were acquired using two BrainProducts LiveAmp amplifiers at a sampling rate of 500~Hz. Electrode impedances were maintained below 50~k$\Omega$ throughout the recording session. FCz served as the online reference electrode, and AFz was used as the ground. 

\subsection{ERP Preprocessing \& Analysis}

EEG preprocessing was performed using MNE-Python and followed established pipelines for ERP research. First, we automatically identified and interpolated bad or outlier channels using the Random Sample Consensus (RANSAC) method~\cite{bigdely2015prep}, employing spherical spline interpolation based on Perrin’s algorithm~\cite{perrin1989spherical}. A notch filter at 50~Hz was applied to remove line noise, and the signal was band-pass filtered between 1--15~Hz to isolate the frequency range relevant for cognitive ERP components. Data were then re-referenced to the common average reference (CAR).

Artifact correction was performed using Independent Component Analysis (ICA) with the extended Infomax algorithm~\cite{lee1999independent}. Components corresponding to eye blinks, eye movements, and muscle noise were automatically identified and removed using the ``ICLabel'' plugin integrated with MNE~\cite{li2022mne}.

The continuous EEG signal was segmented into epochs ranging from 200~ms before to 1000~ms after stimulus onset. A 200~ms pre-stimulus baseline was applied for baseline correction. We focused on three ERP components of interest: N2, P3 and the LPP. 

For the N2, we quantified mean amplitude in the 200--350ms time window, comprising the electrodes Fz, Cz, FC1, FC2 \cite{n2p3, n2_blind, Hong2017Segregating}.
For P3, mean amplitude was quantified in the 300--500 ms window for the P3 and the 400--800 ms window for the LPP \cite{polich2007, CODISPOTI2006213}, time-locked to image onset.
Based on prior literature~\cite{HAJCAK_emotion, STRUBER2002187}, we selected a centro-parietal region of interest (ROI) comprising electrodes Pz, CPz, Cz, CP1, CP2, P3 and P4 for P3 and  Pz, CPz, Cz, CP1, CP2, P1 and P2 for LPP effects are typically observed in response to emotional and source-related facial information. These components were extracted for statistical comparison across experimental conditions (Source $\times$ Label) in our statistical modeling, as presented in \autoref{sec:results}.

\subsection{Sample Size Justification \& Participants}

A priori power analysis was conducted using \textit{G*Power} 3.1 \cite{faul2009statistical} to determine the required sample size for a linear-mixed model, repeated-measures with six within-subject conditions. The analysis assumed a medium effect size ($f = .25$) based on HCI guidelines \cite{yatani2016effect}, a significance level of $\alpha = .05$, desired power of $1 - \beta = .95$, a correlation among repeated measures of $r = .5$, and a nonsphericity correction of $\epsilon = 1$. The results indicated that a total sample size of $N = 30$ participants would be sufficient to detect an effect with this design. The resulting actual power was $>.95$.

We recruited participants through institutional mailing lists and convenience sampling. The sample had a mean age of $26$ years ($SD = 5.57$) and included 17 women, 13 men, none identifying as non-binary or gender-diverse. To assess familiarity and confidence with AI-related concepts, all participants completed the Meta AI Literacy Questionnaire~\cite{carolus2023mails}. The overall AI literacy score was moderate ($M = 5.88$, $SD = 1.33$), with subscale means and standard deviations as follows: AI Literacy ($M = 6.20$, $SD = 1.63$), Create AI ($M = 4.19$, $SD = 2.20$), AI Self-Efficacy ($M = 5.27$, $SD =SD = 1.66$), and AI Self-Competency ($M = 6.41$, $SD = SD = 1.16$). The study was approved by the local ethics committee, qualifying as fast-track approval since the participants were not subjected to any risk (e.g., deception, stress beyond normal levels).

\section{Results}
\label{sec:results}

All analyses were performed at the trial level, so that within-participant
variability was modeled rather than averaged away. Electrophysiological and
eye-tracking data were analyzed within a single Bayesian hierarchical framework
implemented in \texttt{brms} \cite{burkner2017brms}, which estimates group- and
individual-level parameters jointly and returns a full posterior for every effect
rather than a decision against a threshold
\cite{kay2016, schad2021, vandeschoot2021}. This suits the present design for two
reasons. Our central claim concerns an absence of difference between AI- and
human-sourced faces, and a posterior concentrated near zero states that directly,
whereas a non-significant test does not distinguish absence of an effect from
absence of evidence \cite{dix2022}. The hierarchical structure also lets
participants and stimuli be treated as sampled rather than fixed, so estimates
are not tied to the particular faces and people we happened to observe
\cite{schad2021}.

All models shared the same random-effects structure,
\texttt{(1 + Source * Label | PId) + (1 | Image)}, giving participant-specific
intercepts and slopes for the experimental factors together with by-image
intercepts. Likelihoods were adapted to each outcome. Single-trial ERP amplitudes,
averaged across the component ROI, were modeled with a Student-$t$ likelihood to
accommodate the heavy tails typical of electrophysiological data
\cite{alday2021away}. Fixation counts were modeled with a negative binomial
likelihood; fixation durations and dwell time were log-transformed and modeled
with a Student-$t$ likelihood; first-fixation latency was modeled with a Gaussian
likelihood.

The two modalities differ in how the label factor enters. The ERP models use the
three-level \textsc{LabelCond}, whose crossing with \textsc{Source} defines
congruence: a label either matches the face (AI face, ``Made with AI''; human
face, ``Made by a human'') or contradicts it. We report this interaction as
congruence contrasts, congruent and incongruent, each relative to unlabelled
faces. The eye-tracking models code \textsc{Label} as presence versus absence,
because a label is a visible object added to the display and is expected to affect
exploration through its presence rather than its wording; there the interaction
stays on the \textsc{Source}\,$\times$\,\textsc{Label} scale. \textsc{Source} is
coded with AI-generated faces as the reference level in every model, and the two
modalities use opposite reference levels for the label factor: the ERP contrasts
are relative to unlabelled faces, whereas the eye-tracking contrasts are relative
to labelled trials. We report all effects in the text in their substantive
direction. Where an interaction is credible, the corresponding main effect is
conditional on the reference level of the other factor rather than averaged across
it.

\subsection{ERPs}
\label{sec:erp_stats}

\begin{figure*}[h]
  \centering
  \includegraphics[width=\columnwidth]{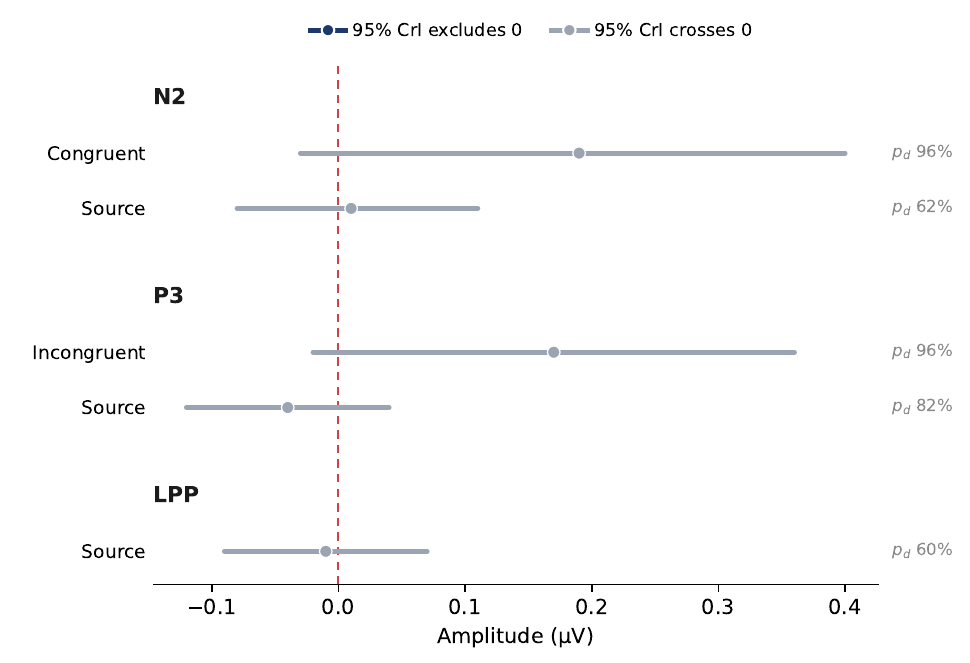}
  \caption{Posterior effects on the three ERP components. Each row is a
  contrast: the marginal difference between AI- and human-sourced faces
  (\textsc{Source}), and the congruent and incongruent labelling conditions,
  each relative to unlabelled faces. Points are posterior medians and bars 95\%
  credible intervals; a row is drawn dark when its interval excludes zero and
  grey when it crosses zero, with the probability of direction ($p_d$) at right.
  Source sits close to zero in every component, while the label-related movement
  appears in the congruence contrasts: the N2 grows with a label present and the
  P3 with label--face conflict, whereas the LPP is unmoved by either factor.}
  \label{fig:erp-forest}
\end{figure*}

\begin{figure*}[t]
  \centering
  \includegraphics[width=\textwidth]{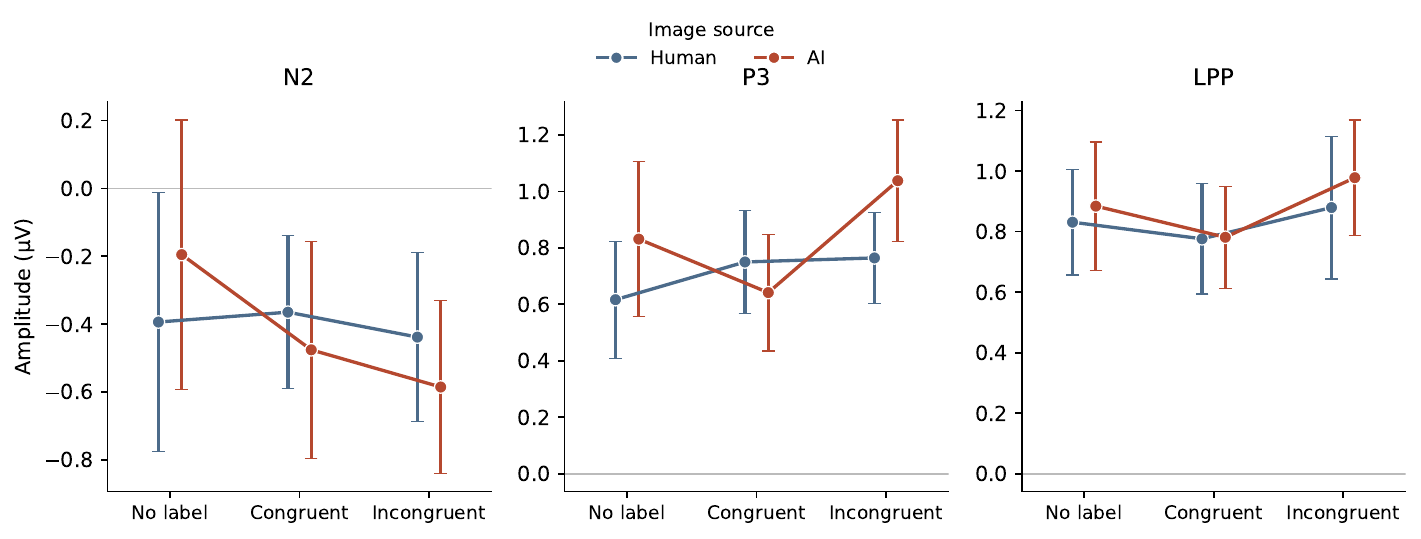}
  \caption{Observed amplitude means with 95\% within-subject confidence
  intervals (Cousineau--Morey), by image source and labelling condition, for
  each component; amplitudes are averaged over the component region of interest
  (frontocentral for the N2, centroparietal for the P3 and LPP). The AI and
  human traces overlap at every level, showing that actual source did not
  separate the response. The condition axis carries what movement there is: the
  N2 becomes more negative once a label is present, the P3 rises under
  incongruence and most so for AI faces, and the LPP stays flat.}
  \label{fig:erp-condmeans}
\end{figure*}

\subsubsection{N2}

Actual image source did not modulate the N2: AI- and human-sourced faces
elicited equivalent frontocentral amplitudes ($Mdn = 0.01~\mu$V, 95\% CrI $[-0.08, 0.11]$, $p_d = 61.5\%$). What moved the early response was the label rather than the face: relative to unlabeled faces, labeled faces tended to elicit a larger N2, most clearly when label and face were congruent ($Mdn = 0.19~\mu$V, 95\% CrI $[-0.03, 0.40]$, $p_d = 95.7\%$). The N2 thus appears sensitive to the presence of a source label rather than to label--face conflict per se.

\begin{figure*}[t]
        \centering
        \includegraphics[width=\textwidth]{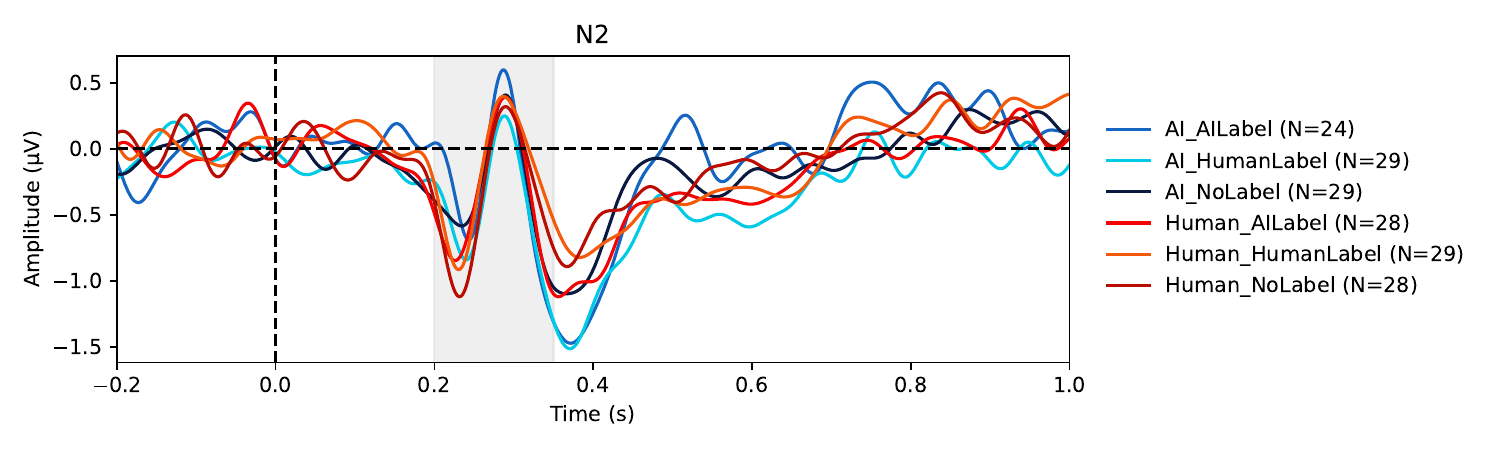}
        \caption{\textbf{N2 grand average waveform.} Averages for the six conditions (AI/Human
face $\times$ Made with AI/Made by a human/No label), measured over fronto-central
electrodes (Fz, Cz, FC1, FC2); the shaded region (200--350\,ms) marks the analysis window.
AI and human faces produced the same response. Any face carrying a label, whether the
label matched the face or not, produced a larger response than a face with no label at
all, though the difference is small. The earliest sensitivity is therefore to the presence
of a source cue, not to its agreement with the face, which is evaluated at a later stage.}
\label{fig:n2}
\end{figure*}
\subsubsection{P3}

The P3 likewise showed no difference by actual source ($Mdn = -0.04~\mu$V, 95\%CrI $[-0.12, 0.04]$, $p_d = 81.7\%$). Instead, it was here that label--face \emph{conflict} registered: incongruent trials---an AI face tagged ``Human,'' or the reverse---produced a larger P3 than unlabelled faces ($Mdn = 0.17~\mu$V, 95\% CrI $[-0.02, 0.36]$, $p_d = 95.6\%$), consistent with the P3's role in stimulus re-evaluation. The mismatch between what the face is and what the label claims, not the face's true origin, drove this later response.

\begin{figure}[t]
        \centering
        \includegraphics[width=\textwidth]{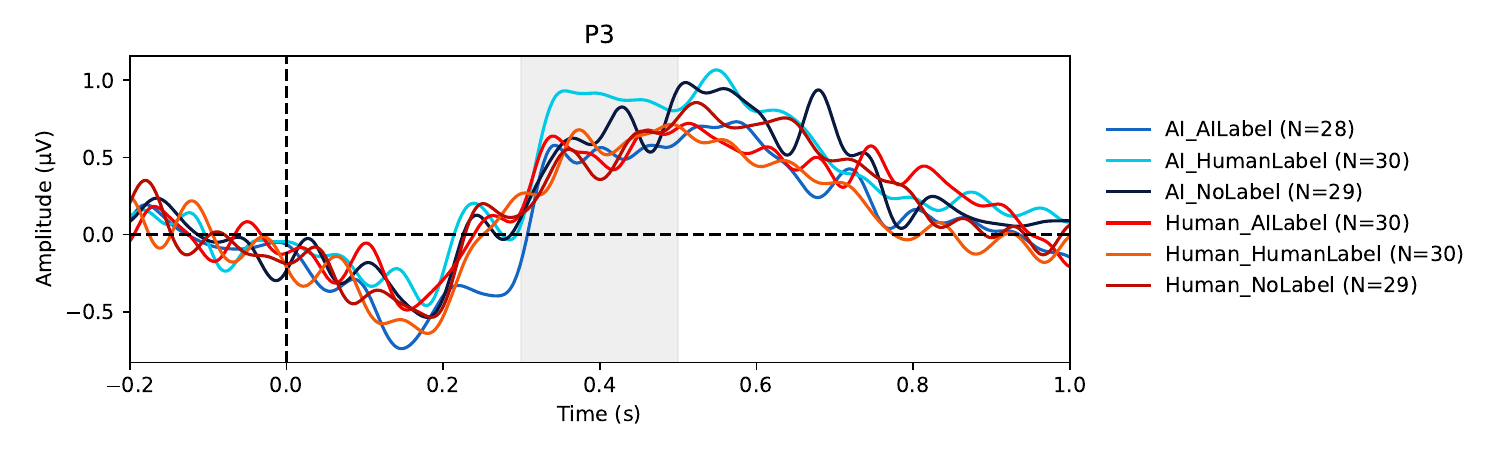}
        \caption{\textbf{P3 grand average waveform.} Averages for the six conditions (AI/Human
face $\times$ Made with AI/Made by a human/No label), measured over centro-parietal
electrodes (Pz, CPz, Cz, CP1, CP2, P3, P4); the shaded region (300--500\,ms) marks the
analysis window. AI and human faces produced the same response. An AI face labelled
``Made by a human'', or a human face labelled ``Made with AI'', produced a larger response
than a face with no label at all.}

        \label{fig:p3}
\end{figure}

\subsubsection{LPP}

The affective window was unaffected by both factors. Neither actual source ($Mdn = -0.01~\mu$V, 95\% CrI $[-0.09, 0.07]$, $p_d = 60.0\%$) nor congruence (all $|Mdn| \leq 0.08~\mu$V) modulated the LPP,indicating that labelling effects did not extend into sustained affective processing. 

\begin{figure}[t]
  \centering
  \includegraphics[width=\textwidth]{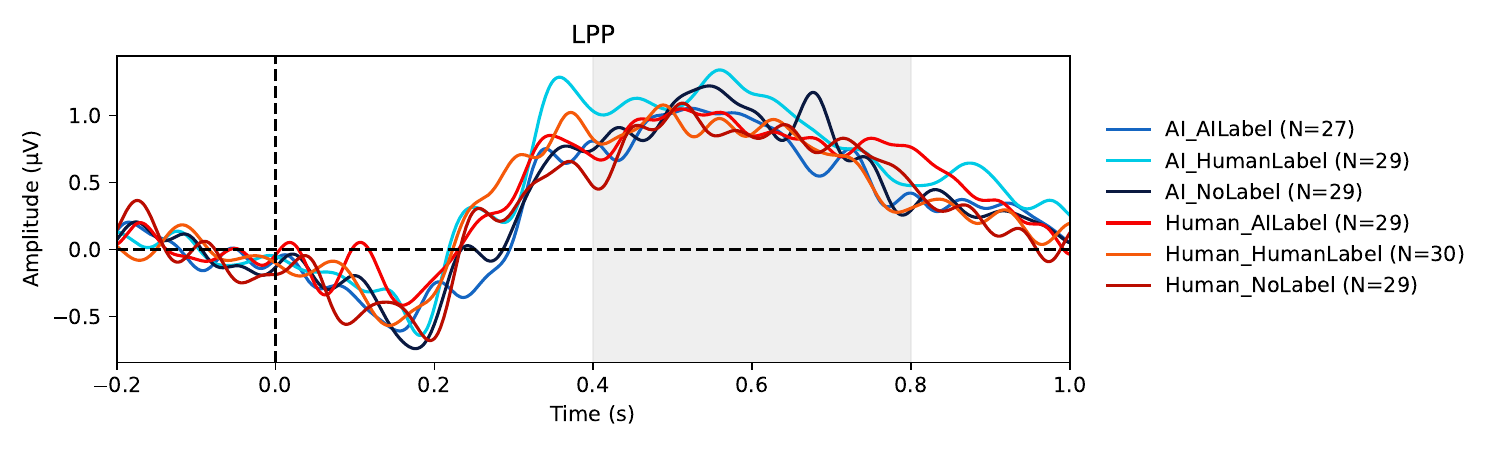}
  \caption{\textbf{LPP grand average waveform.} Stimulus-locked averages for the six
experimental conditions (AI/Human source $\times$ Made with AI/Made by a human/No label),
computed over centro-parietal electrodes (Pz, CPz, Cz, CP1, CP2, P1, P2). The shaded
region (400--800\,ms) indicates the LPP analysis window. A sustained positivity rises from
roughly 200\,ms and is maintained across the window in every condition. The six traces
overlap throughout, and the models estimate no difference by actual source
or by label congruence. Whatever a disclosure label did to earlier processing did
not carry into this later affective stage.}
  \label{fig:lpp}
\end{figure}
\section{Eye-Tracking}
\begin{figure}[!t]
  \centering
  \includegraphics[width=0.7\columnwidth]{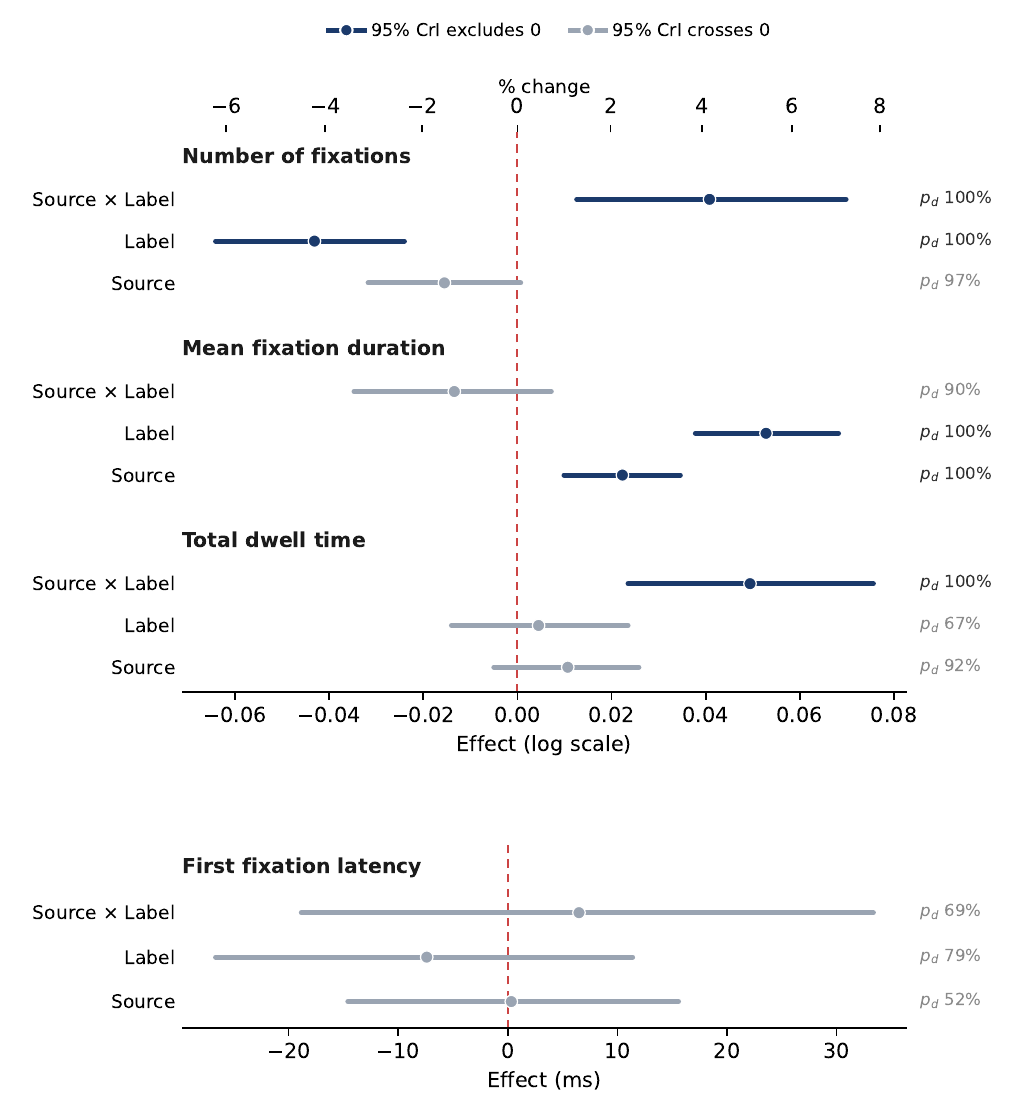}
  \caption{Posterior effects of image source, disclosure label, and their
  interaction on each eye-tracking measure. Points are posterior medians and
  bars the 95\% CrIs; the upper axis restates the log-scale coefficients as
  percentage change. Effects whose interval excludes zero are dark, those
  overlapping zero are grey, and $p_d$ is given at right. 
  Labels credibly shifted how faces were explored while leaving the speed of
  initial orienting unchanged, and image source registered only in fixation
  duration.}
  \label{fig:gaze-forest}
\end{figure}

\begin{figure*}[!t]
  \centering
  \includegraphics[width=\textwidth]{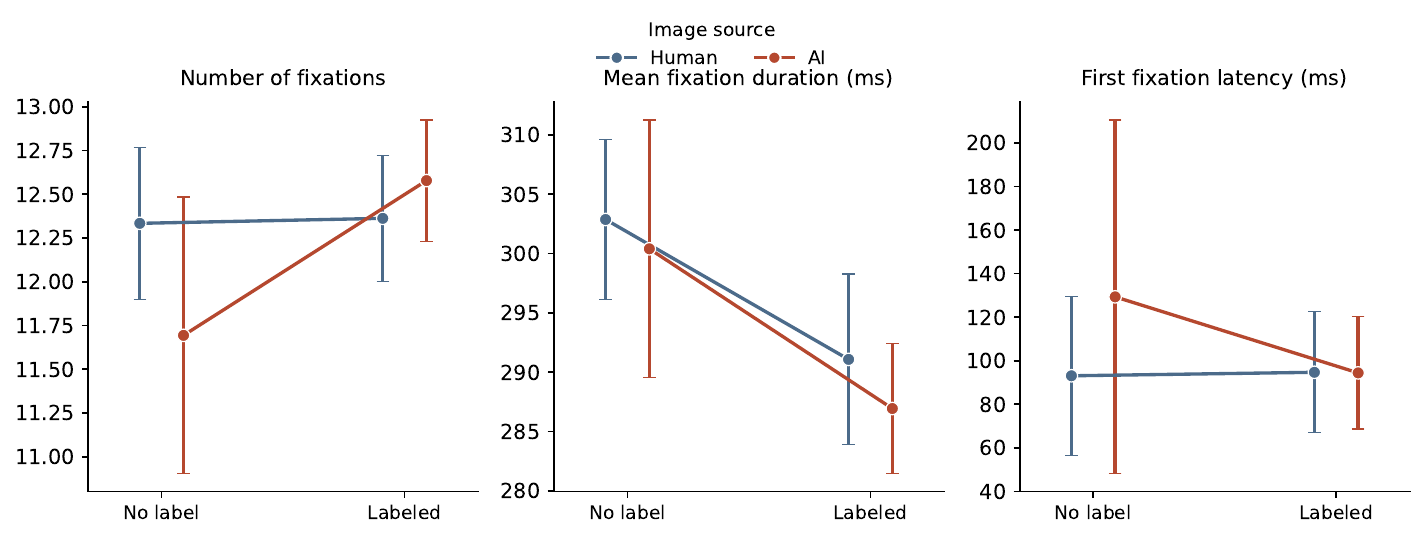}
  \caption{Observed cell means with 95\% within-subject confidence intervals
  (Cousineau--Morey) for each measure, by image source and label presence, on
  the natural scale of each measure; total dwell time is omitted as redundant.
  The panels show the pattern behind the modelled effects: fixations were
  shorter and more numerous when a label was present, the fixation-count shift
  fell almost entirely on AI faces, and the timing of the first fixation
  overlapped across all conditions.}
  \label{fig:gaze-condmeans}
\end{figure*}




\subsection{Number of Fixations}
Labeled faces were fixated more often than unlabeled ones ($Mdn = -.043$, 95\% CrI
$[-.064, -.024]$, $p_d = 100\%$). These extra fixations were shorter (see mean fixation duration below), so participants looked at the face more times without spending longer on it overall. The increase was larger for AI-generated faces than for human faces ($Mdn = .041$, 95\% CrI $[.013, .070]$, $p_d = 99.8\%$).
Actual source on its own had a smaller and less certain effect. AI faces were fixated slightly more often than human faces, but the credible interval includes zero ($Mdn = -.016$, 95\% CrI $[-.032, 0.001]$, $p_d = 97.1\%$).

\subsection{Mean Fixation Duration}
Mean fixation duration carried credible effects of both source ($Mdn=.022$,
95\% CrI $[.010,.035]$, $p_d=100\%$) and label ($Mdn=.053$, 95\% CrI
$[.038,.068]$, $p_d=100\%$). Fixations were shorter on AI faces than on human
faces, and shorter again when a label was present. The Source$\times$Label interaction was not credible ($Mdn=-.013$, 95\% CrI $[-.035,.007]$, $p_d=89.9\%$), so labeling shortened fixations by a comparable amount for both sources.

\subsection{Total Dwell Time}
Total dwell time is the product of fixation count and mean fixation duration,
and in these data the three are effectively collinear, so its coefficients
follow from the two measures above rather than adding independent evidence; we
report them for completeness. Neither source ($Mdn=.011$, 95\% CrI
$[-.005,.026]$, $p_d=91.6\%$) nor label presence ($Mdn=.005$, 95\% CrI
$[-.014, .024]$, $p_d=67.1\%$) produced a credible difference on its own, and
the credible Source$\times$Label interaction ($Mdn=.049$, 95\% CrI
$[.024,.076]$, $p_d=100\%$) restates the fixation-count interaction already
reported.

\subsection{First Fixation Latency}
The latency of the first fixation gave no credible effect of source
($Mdn=0.316$, 95\% CrI $[-14.593,15.573]$, $p_d=51.8\%$), label ($Mdn=-7.397$,
95\% CrI $[-26.664,11.384]$, $p_d=78.6\%$), or their interaction ($Mdn=6.494$,
95\% CrI $[-18.840,33.353]$, $p_d=69.0\%$). 

\section{Discussion}
Whether a disclosure label changes how a face is processed, or only what a viewer reports afterwards, has been hard to establish because the two causes travel together: work that varies what an image is leaves belief about its origin to chance \cite{nightingale2017, miller2023ai, schindler2017}, and work that varies belief does so over a single set of real photographs \cite{eiserbeck2023deepfake, conrad2025truthfulness}. Study~1 fixed the perceptual status of the stimuli in advance, so that any later effect of a label could not be traced to visual ambiguity. Study~2 crossed source against the label attached to each face, recording EEG and gaze during stimulus presentation. The presence of a label changed early processing; a label that contradicted the face, altered a later, evaluative stage, and labels reshaped how faces were visually explored. Actual source of the face did not separate the neural response. Together, the two studies contribute a validated stimulus set and evidence that the origin attributed to a face affects cognitive processing, unlike its true origin. 
In the following we summarize the results and discuss them in light of our research questions.

\subsection{Summary of Results}

The ERP results separated what a face is from what it is labeled as. At the N2, P3, and LPP, AI- and human-sourced faces produced amplitudes that did not differ credibly, with every posterior centered near zero, so actual source left no credible mark on stimulus-locked processing. Instead, the label carried the found effect. A present label enlarged the N2 relative to unlabeled faces whether or not it matched the face, which places the earliest sensitivity at the presence of a source cue. Label--face incongruence caused a stronger P3 response, in line with that component's role in stimulus re-evaluation \cite{polich2007}. The LPP moved with neither factor.

The gaze results matched this shape at the level of overt attention. The latency of the first fixation showed no credible effect of source, label, or their combination, so we have no evidence that initial orienting shifted. What labels changed was the later exploration of a face: a labeled face drew more fixations than an unlabeled one and each fixation was shorter, and the rise in fixation count fell largely on AI-generated faces. Actual source surfaced once, in slightly shorter fixations on AI faces, and it reached neither the EEG response nor the first fixation.

Across measures, the label assigned to a face shaped its processing while its true origin did not register on these measures. The effect appeared after a face had been detected and fixated, in the re-evaluative and exploratory measures rather than in early detection or the first capture of attention, which locates AI disclosure on the evaluation of a face rather than on its perception. 





\subsection{Electrophysiological Signatures of Source and Label}
\paragraph{Actual source left no credible ERP proxy.}

In \textbf{RQ1} we asked whether a face's actual source, AI-generated or human, modulates its electrophysiological processing at the early attentional (N2), evaluative (P3), and sustained affective (LPP) stages. Actual source left no measurable trace on any of the three components. AI- and human-sourced faces produced equivalent amplitudes, with each posterior centered on zero. Under our conditions, the EEG response did not discriminate whether a face was real or synthetic.

Previous ERP studies have reported differences between real and artificial faces. Schindler et al.~\cite{schindler2017}, for example, varied face realism from cartoon to photorealistic images and found differences in both the N170 and LPP. Similarly, Wheatley et al.~\cite{wheatley2011} compared human faces, doll faces, and objects. They found similar early responses to human and doll faces, but only human faces produced a sustained late positivity after 400~ms, which they associated with the perception of another mind. Based on these findings, we might therefore expect the actual source of a face to affect at least the later ERP components.

We did not find such an effect. One explanation is the difference in the visual appearance of the stimuli. The earlier studies compared faces that were clearly different in appearance, such as cartoons and photographs or dolls and human faces. In our study, both AI-generated and human faces were photorealistic. This is important because previous work suggests that the distinction between artificial and real faces becomes less pronounced when artificial faces are highly realistic. Geiger and Balas~\cite{geiger2021robot}, for example, found that realistic artificial human faces produced N170 responses similar to those elicited by real faces. Our findings extend this observation to diffusion-generated faces and to later stages of processing. Under our experimental conditions, the actual source of the face did not produce a clear difference in the N2, P3, or LPP.

The difference from ~\citet{wheatley2011} may be particularly relevant for the LPP. Their late positivity separated human faces from dolls, which were visibly inanimate. In contrast, our AI-generated faces were photorealistic and depicted people, so both the AI and human faces provided similar visual cues of another person. This could explain why the LPP did not differ between the two sources in our study. We consider this a possible explanation rather than a direct test, since the AI and human faces came from different databases and were not matched on low-level visual features. Still, we found no evidence of a source effect despite these differences.
Our design also allows us to address a limitation of previous work. Studies that manipulate realism change both the appearance and the source of the face, while studies that manipulate beliefs about origin keep the actual source constant. We controlled the perceived origin in Study~1 and crossed actual source with the label in Study~2, thus allowing to separate these factors. Our results suggest that when AI-generated faces are sufficiently realistic, their actual source does not produce a measurable difference in stimulus-locked ERP responses. Instead, the response appears to depend more on the visual appearance of the face and on the information provided by the label. These findings are in line with \citet{lago2021more}'s work, which demonstrated that generative models have advanced in producing synthetic faces that are increasingly difficult to be distinguished from real ones. 

For HCI, this means that viewers may not have a perceptual signal that tells them that a realistic face is AI-generated. If users need to know the origin of an image, this information therefore needs to be provided on the interface, either through provenance mechanisms or AI disclosures. Disclosure labels can convey this information directly, without relying on users to infer the source from the face itself.

\paragraph{How Source Labels Shape Responses to AI Content}

In \textbf{RQ2}, we asked whether people responded to the label itself and to whether the label matched the face. We found effects in both the N2 and P3, but the two components reflected different parts of the response. The N2 was larger whenever a label was present, whether or not it matched the face. The P3 was larger when the label and the source of the face did not match. This suggests a simple two-stage process: people first noticed the label, and then they responded when the label conflicted with the face.

This result differed from our initial expectation that the N2 would mainly reflect a conflict between the label and the face. A simpler interpretation is that the N2 reflects the early registration of a task-relevant cue \citet{n2p3}. This is consistent with Folstein and Van Petten~\cite{folstein2008}, who distinguish between N2 responses related to cognitive control or salient events and those more specifically related to mismatch. The P3 then reflected the later mismatch, consistent with its role in updating an interpretation when new information does not fit the current one~\cite{polich2007}. In practical terms, the label may be noticed first and compared with the face only afterward.

This sequence is also consistent with the MAIN model~\cite{sundar2008, sundar2019}: In an interface, a cue such as an “AI-generated” label can activate an expectation about the content's origin before the user examines it in detail, leaving that expectation to be checked against the content itself. Following this interpretation, the N2 should reflect noticing the source cue, while the P3 reflects the later conflict when the cue and the face do not fit together. We use the model to describe how the cue may be processed, not to claim that labels necessarily increase or decrease trust, which our study did not test.

Moreover, our findings extend the work of ~\citet{conrad2025truthfulness}, who found that an AI warning affected trust judgments while participants viewed AI-generated political images. In their study, all images were AI-generated, so the warning could only be present or absent. It could not be correct for some images and incorrect for others. Our design allowed us to create this mismatch and separate a response to seeing a label from a response to the label being wrong. This gives a clearer picture of how people may process source information in interfaces that include AI labels.

Our findings have implications for HCI and human--AI interaction: a label can affect how users process content as soon as they see it, before they have fully evaluated the content itself. For example, an “AI-generated” label on a social media post, image search result, or decision-support interface may shape the user's initial expectations. Labels should therefore be accurate, clear, and used only when they provide useful information. A small cue may be enough to make the source noticeable, but an incorrect or unnecessary label could lead users to interpret content as machine-generated when that interpretation is not warranted.

\paragraph{The LPP boundary separates evaluation from affect.}
In our findings, neither source nor label congruence modulated the LPP, indicating that the disclosure effect does not extend into affective processing. This result stands in contrast to \citet{eiserbeck2023deepfake}, who reported late-positivity effects when participants believed a smiling face was fake. That effect is likely attributable to the used stimuli, which confounded belief with appearance: smiling adds affective content that cannot be separated from the source. Our study design avoids this confound, since appearance was held constant and the label carried no affective content - we can directly attribute the absence of an LPP effect to disclosure of the label. 
These results build a positive case for disclosure labeling, supporting \citet{susser_2018}'s distinction between legitimate persuasion and emotional manipulation: the labels appeal openly to a person's deliberative capacities by offering information to weigh, rather than manipulating the emotional state of the user. Here, the absence of LPP modulation indicates that the label effects were tied to evaluative processing - the categorization and re-evaluation reflected in the N2 and P3, indicating that the label informs the viewer's judgment of the artifact rather than exploiting affect.




\subsection{Gaze Signatures of Source and Label}
In this section we turn from brain processing to visual behavior. Addressing whether disclosure labels change how a face is visually explored - and whether this depends on the face's actual source (\textbf{RQ3}), as well as asking whether the image source or label shape the earliest allocation of attention (\textbf{RQ4}). 

\paragraph{Labels changed how a face was explored}

\textbf{RQ3} asked whether disclosure labels change how a face is visually explored, and whether this depends on its actual source. We found that labeled faces received more fixations, but these fixations were shorter, while total viewing time remained unchanged. Participants therefore distributed roughly the same viewing time across more locations. We interpret this as a change in visual exploration rather than in the overall time spent viewing the face. This interpretation is necessarily cautious, since fixation count and duration alone cannot establish a shift between ambient and focal viewing, which also requires saccade amplitude \cite{krejtz2016coefficientk}.

The effect of labeling was larger for AI-generated than for human faces. Thus, the effect of the label on viewing depended on the face it was attached to, even though EEG showed no clear source effect. One possible explanation is that viewers used the label as a cue to look back at the face and assess whether it fit the stated origin. The extra fixations may reflect this additional inspection, although our data cannot show that directly. This would place the source-related gaze effect later than the EEG effects: after the label has been registered, viewers may return to the image and explore it further. This is consistent with the different timing of the effects across the two measures.

This finding provides a partial dissociation between neural and gaze responses. Actual source was not reflected in stimulus-locked EEG, but it did matter for how the faces were explored. We treat this interaction as provisional, since the AI and human face sets came from different databases and were not matched on low-level visual properties \cite{resulbegoviq2026eyetracking}.

Our findings are consistent with previous eye-tracking work suggesting that source effects depend on the task. \citet{bozkir2024can}, for example, found no reliable difference in free-viewing time between real and GAN-generated faces, with task phase and priming having a larger effect than actual origin. Our results extend this finding to diffusion-generated faces in a setting where source information was also shown on screen. Other studies have found differences between authentic and manipulated content \cite{gupta2020fakeet, cartella2024unveiling}, but these studies changed the content itself and asked participants to detect manipulation. Our task involved static faces under free viewing, where gaze may be less sensitive to differences in source \cite{peterson2012below}.

\textbf{RQ4} addressed a different stage of attention: whether source or labeling affects the latency of the first fixation. We found no clear effect of either factor. This suggests that neither the actual source of the face nor the disclosure label reliably changed how quickly participants first oriented to the stimulus. This is consistent with the limited evidence on early orienting to AI-generated content, which has not shown that generated material reliably attracts gaze faster than authentic material \cite{resulbegoviq2026eyetracking}.

For HCI, the two RQs point to a distinction between getting attention and shaping attention after viewing has started. Source labels did not clearly change initial orienting, but they did change how participants explored the face. A label is therefore not simply information placed next to an image; it becomes part of the visual interaction. Its size, placement and level of detail may affect how users distribute their limited viewing time between the disclosure and the content itself \cite{gamage2025labeling, prajod_dislosure}. For human--AI interaction, source disclosure should therefore be considered as part of interface design, particularly when labels are presented alongside visual content.

\subsection{ERPs and Gaze Agree on When Disclosure Takes Effect}
Across \textbf{RQ2}--\textbf{RQ4}, the EEG and eye-tracking results point to a similar pattern: disclosure did not clearly change the earliest stage of viewing, but its effects emerged once the face was being processed and explored. For \textbf{RQ4}, neither source nor label affected the latency of the first fixation. For \textbf{RQ1}, actual source also did not affect any of the three ERP components. The effects associated with labeling appeared later: the P3 was larger when the label conflicted with the face's actual source, while gaze showed more fixations and shorter fixation durations for labeled faces. Together, these findings suggest that disclosure does not determine whether a face attracts attention or how quickly it is first seen; rather, its effects are more apparent in what happens \textit{after} viewing has started.

Using both measures also helps distinguish when these effects occur. The ERP analysis captures activity during the first second after image onset, whereas the eye-tracking measures cover the full presentation. The gaze results therefore suggest that the effect of disclosure can continue after the initial evaluative period captured by the ERP analysis, with labeled faces receiving more, shorter fixations. This complements the P3 finding. Previous work has combined EEG and gaze to study responses to synthetic media. \citet{gupta2020fakeet}, for example, found differences between real and manipulated videos in both measures, although their participants were explicitly asked to detect manipulation and viewed dynamic content. These task and stimulus differences may explain why actual source registered in their study but not consistently in ours. More broadly, \citet{Sun2026ailabelshealthtrust} show how behavioral and physiological measures can capture different parts of a response to AI-generated information. Co-registration of gaze and EEG has also been used in HCI to relate visual attention to stimulus-locked activity \cite{chiossi2024understanding}. Our findings similarly suggest that disclosure effects are easier to understand when measures capture different stages of the interaction.

The two modalities, however, do not tell exactly the same story: actual source appeared in the gaze data through shorter fixations on AI faces and a larger label effect for AI-generated faces, but it did not appear in the EEG. Thus, source may affect how viewers explore an image without producing a corresponding difference in stimulus-locked brain activity. We treat this dissociation cautiously because the AI and human face sets came from different databases and were not matched on low-level visual properties \cite{resulbegoviq2026eyetracking}. The neural and gaze findings therefore provide a partial rather than complete separation between actual source and disclosure.

For HCI, this distinction matters because disclosure can affect the interaction after a user has already looked at the content. A design may leave initial attention unchanged while still changing how users inspect the content afterwards. This is difficult to capture with post-exposure ratings alone, which measure the outcome of the evaluation rather than how it unfolded \cite{conrad2025truthfulness, jakesch2019}. Combining behavioral measures such as gaze with physiological measures can therefore provide a more detailed account of how disclosure works (cf., \citet{Sun2026ailabelshealthtrust}). In practice, gaze is also considerably easier to deploy than EEG, making it a more feasible tool for evaluating source-disclosure designs in human--AI interaction.

\subsection{Designing Disclosure Labels}

Disclosure labels are usually treated as information that viewers can consider alongside what they see. Our results suggest that this is not always how they work. For the faces in our study, participants showed no clear sensitivity to actual source in the ERP measures, in how quickly they looked at the faces, or in how many fixations they made. We therefore found no evidence that they could reliably determine from the face itself whether it was AI-generated or human-made. The label provided information about the face's source that could not be reliably determined from the face itself.

In other settings, however, provenance may not be binary. Content can involve different degrees and types of human and AI contribution, and the way this information is disclosed can affect how users understand that contribution \cite{kusters2026more}. This makes clear and accurate provenance information important, particularly when viewers cannot independently assess how the content was produced. If viewers can judge the source independently, they may be able to reject an incorrect label. If they cannot, there is little they can use to do so. The accuracy of the labeling process therefore becomes especially important, which supports using verifiable provenance rather than relying on post-hoc classification \cite{feng2023examining}.

Our design also tested a case that current regulation does not require. The EU AI Act and comparable rules require providers to disclose synthetic origin \cite{Elali2024aidisclosures, li2026governing}, so AI-generated content is marked while human-made content is generally left unmarked. We labeled both human and AI-generated faces. The early ERP response was larger whenever a label was present, with no clear difference between the two label types. Participants therefore responded to the presence of a source cue before there was a clear response to what the cue said. In an asymmetric system, the absence of a label does not provide the same cue. People can still draw an inference from that absence: \citet{pennycook2020} found that unwarned content is often treated as verified and rated as more accurate. With synthetic faces, there is little opportunity to correct that inference from the image itself. Labeling both origins, as we did, could reduce this ambiguity, but the way provenance is presented may also shape how users interpret the relative human and AI contribution \cite{kusters2026more}. Whether a positive human-made label can also reduce the confidence attached to otherwise unlabeled content remains to be tested.

The incongruent condition gives us a small test of what happens when the disclosure is wrong. When the label contradicted the face's actual source, the response was more consistent with the label than with the actual source, and we found no clear indication that participants recovered the true origin. A missing or incorrect label is therefore not necessarily just missing information. In our data, an incorrect label provided information that viewers had no reliable way to reject. This is relevant for systems that assign labels automatically, since errors will occur and viewers may not be able to detect them. Reporting labeling accuracy, and where possible communicating uncertainty in the attribution, may therefore matter more than refining the wording of the label itself.

Placement also has a cost. Our label appeared in the bottom-right corner, following \citet{gamage2025labeling}, far enough from the center that reading it required a saccade. Labeled faces received more and shorter fixations without an increase in total viewing time. The label therefore changed how participants distributed their viewing time rather than increasing the time they spent with the image. Where a disclosure is placed is consequently also a decision about where users spend their attention. This may matter more in feeds and search results, where people spend less time with individual items than the four seconds allowed in our study.


\section{Limitations \& Future Work}

A few limitations should be kept in mind when interpreting this work.
The main limitation is that our two stimulus sets were not matched on low-level visual properties. Human faces came from the Chicago Face Database \cite{Ma2021Chicago}, while AI faces came from a standardized Stable Diffusion collection. Both sets were standardized differently, with the photographs using a white background and gray clothing and the AI images showing their own regularities, such as pastel backgrounds and smoothed skin. This provides an alternative explanation for the shorter fixations on AI faces and the larger label effect for them. \citet{miller2023ai} also report that perceived AI-likeness can vary with apparent race, which we did not control. Future work should therefore match AI and human faces on visual properties and apparent demographics. At the same time, tighter matching would move the stimuli further from the varied faces people encounter on social platforms and online news \cite{yang2024characteristics,peng2025crafting}. Future studies should therefore also test more naturalistic images across different settings and generators, allowing source effects to be separated from stimulus appearance without sacrificing ecological validity.

Study 1 also fixed perceived origin by selecting faces that raters reliably classified as AI-generated. This removed ambiguous cases but also meant that perceived and actual origin were not fully crossed. Future work should cross perceived origin, actual origin, and label to test whether labels still guide responses when an AI-generated face looks convincingly human.

Our participant sample was small, relatively young, and reported moderate AI literacy. It did not include older adults or people with less experience with synthetic media, although attention to credibility cues can vary with digital literacy \cite{steinfeld2023credible}. A more diverse sample is therefore needed.

We also tested static faces under free viewing, with one label design in one position. Real disclosures appear in feeds and search results, where viewing times are shorter and labels compete with surrounding content. They also vary in wording, detail, and visual form \cite{gamage2025labeling}. Future work should test these variations, as well as content with visible generation artifacts, where labels supplement rather than replace perceptual cues.





\section{Conclusion}

We separated actual source from attributed origin to test whether disclosure labels change how AI-generated and human faces are processed. Across an online validation study ($N = 169$) and a lab study ($N = 30$), actual source did not produce clear differences in N2, P3, or LPP amplitude, or in how quickly faces were first fixated. Labels, in contrast, changed how faces were explored: labeled faces received more, shorter fixations, while the ERP results suggest that labels were registered and later compared with the face's source.

These findings make three contributions. First, we provide a perceptually validated set of diffusion-generated faces for separating image source from perceived origin. Second, we show that attributed origin can affect cognitive processing even when actual origin does not, using EEG and eye tracking rather than post-exposure judgments. Third, the findings have implications for disclosure design. When content provides no reliable perceptual cue to its origin, the label becomes an important source of that information, making labeling accuracy critical. Labeling both human and AI content may also reduce the ambiguity created by an unlabeled item, while label placement affects how users distribute their limited viewing time.

\section*{Open Science \& Use of Generative AI}

To support transparency and further research, we have made our full experimental setup, datasets, and analysis scripts openly available on the Open Science Framework: \url{https://osf.io/zejp3/overview?view_only=31ead9b2ec6c4ca88b8fe2d14c62c520}. We encourage others to reproduce, verify, and build upon our findings.
We used a large language model (Claude Opus 5 and Claude Fable 5, Anthropic; z-ai/glm-5.2, Gemini Nanobanana 2.0, and moonshotai/kimi-k2.6) to assist with copy-editing the manuscript, teaser figure design, and with writing plotting and analysis code. All study design, data collection, statistical analysis, and interpretation were carried out by the authors, who take full responsibility for the content.
%
%
%
\begin{acks}

\end{acks}

\bibliographystyle{ACM-Reference-Format}
\bibliography{bibliography}


%
%
\end{document}